\documentclass[a4paper,11pt]{article}
\pdfoutput=1 

\usepackage{jheppub} 

\usepackage[T1]{fontenc} 
\usepackage{subfigure}
\usepackage{hyperref}
\usepackage{silence}
\title{\boldmath Flavor-changing decay $h\to \tau\mu$ at super hadron colliders}

\author[a]{M. A. Arroyo-Ure\~na,}
\author[b]{T. A. Valencia-P\'erez,}
\author[a]{R. Gait\'an,}
\author[a]{J. H. Montes de Oca Y.,}
\author[b] {A. Fern\'andez-T\'ellez.}

\affiliation[a]{ Departamento  de  F\'isica,  FES-Cuautitl\'an,
Universidad  Nacional  Aut\'onoma  de  M\'exico,
C.P.  54770,  Estado  de  M\'exico,  M\'exico.}
\affiliation[b]{Facultad de Ciencias F\'isico-Matem\'aticas\\
Benem\'erita Universidad Aut\'onoma de Puebla, C.P. 72570, Puebla, Pue., M\'exico.}

\emailAdd{marcofis@yahoo.com.mx}
\emailAdd{antonio.valenciap@alumno.buap.mx}
\emailAdd{rgaitan@unam.mx}
\emailAdd{josehalim@comunidad.unam.mx}
\emailAdd{afernand@fcfm.buap.mx}

\abstract{We study the flavor-changing decay $h\to \tau \mu$ with $\tau=\tau^-+\tau^+$ and $\mu=\mu^-+\mu^+$ of a Higgs boson at future hadron colliders, namely: a) High Luminosity Large Hadron Collider, b) High Energy Large Hadron Collider and c) Future hadron-hadron Circular Collider. The theoretical framework adopted is the Two-Higgs-Doublet Model type III. The free model parameters involved in the calculation are constrained through Higgs boson data, Lepton Flavor Violating processes and the muon anomalous magnetic dipole moment; later they are used to analyze the branching ratio of the decay $h\to\tau\mu$ and to evaluate the $gg\to h$ production cross section. We find that at the Large Hadron Collider is not possible to claim for evidence of the decay $h\to\tau\mu$ achieving a signal significance about of $1.46\sigma$ by considering its final integrated luminosity, 300 fb$^{-1}$. More promising results arise at the High Luminosity Large Hadron Collider in which a prediction of 4.6$\sigma$ when an integrated luminosity of 3 ab$^{-1}$ and $\tan\beta=8$ are achieved. Meanwhile, at the High Energy Large Hadron Collider (Future hadron-hadron Circular Collider) a potential discovery could be claimed with a signal significance around $5.04\sigma$ ($5.43\sigma$) for an integrated luminosity of 3 ab$^{-1}$ and $\tan\beta=8$ (5 ab$^{-1}$ and $\tan\beta=4$).}

\begin{document} 

\maketitle
\flushbottom

\section{Introduction}
A lepton flavor violation (LFV) is a transition between $e$, $\mu$, $\tau$ sectors that does not conserve lepton family number. Within the Standard Model (SM) with massless neutrinos, individual lepton number is conserved. Even with the addition of non-zero neutrino masses, processes that violate charged lepton number are suppressed by powers of $m^2_{\nu}/m^2_W$ \cite{raidal} and they should be extremely sensitive to physics beyond the SM (BSM). Neutrino oscillations are a quantum mechanical consequence of the existence of nonzero neutrino masses and mixings. The experiments with solar, atmospheric, reactor and accelerator neutrinos \cite{ahmad, eguchi, fukuda, ahn} have provided evidences for the existence of this phenomenon \cite{pontecorvo, maki}  giving a clear signal of LFV. 
On the other hand, the observation of charged lepton flavor-violating (CLFV) processes  would be a non-trivial signal of physics BSM. However, no evidence of the LFV in the searches of lepton decays 
$\tau^{-} \rightarrow  e^{-} e^{-} e^{+}$ , $\tau^{-} \rightarrow  \mu^{-} \mu^{-} \mu^{+}$ \cite{hayasaka},  and $\mu^{-} \rightarrow  e^{-} e^{-} e^{+}$ \cite{bellgardt},  or radiative decays $\mu \rightarrow  e \gamma$ \cite{baldini}, $\tau \rightarrow  \mu \gamma$, $\tau \rightarrow  e \gamma$  \cite{aubertet} which impose very restrictive bounds on the rates of these processes. Particularly interesting is the decay $h\to\tau\mu$, which was studied first by authors of \cite{LorenzoToscanohtaumu}, with subsequent analysis on the detectability of the signal appearing soon after \cite{TaoHan, KA}. This motivated a plethora of calculations in the framework of several SM extensions, such as theories with massive neutrinos, supersymmetric theories, etc., \cite{LorenzoJHEP, ArgandaCurielHerrero, Brignole, LorenzoMoretti, Goku1, Goku2, LamiRoig, htaumuBUAP}. The observation of the SM Higgs boson with a mass close to 125 GeV at the Large Hadron Collider (LHC) \cite{Aad, Chatrchyan} opened a great opportunity to search for physics BSM, in particular through the decay $h\to\tau\mu$. Currently the upper bounds reported by CMS and ATLAS collaborations \cite{Sirunyan:2019shc, Aad:2019ugc} are
\begin{eqnarray}
\mathcal{BR}(h\to\tau\mu)&<&0.25\%\, (\text{CMS}), \\
\mathcal{BR}(h\to\tau\mu)&<&0.28\%\, (\text{ATLAS}).
\end{eqnarray}
With this values, searches for decay $h\to\tau\mu$ look promising with luminosities larger than the one reached by the LHC ($300$ fb$^{-1}$). This could be achieved at the High Luminosity Large Hadron Collider (HL-LHC) \cite{Apollinari:2017cqg} which will be a new stage of the LHC starting about 2026 with a center-of-mass energy of 14 TeV. The upgrade aims at increasing the integrated luminosity by a factor of ten ($3$ ab$^{-1}$, around year 2035) with respect to the final stage of the LHC. In addition, subsequent searches for the decay $h\to \tau\mu$ could be performed at the High Energy Large Hadron Collider (HE-LHC) \cite{HELHC} and at Future hadron-hadron Circular Collider (FCC-hh) \cite{FCChh}, which will reach an integrated luminosity of up to 12 and 30 ab$^{-1}$ with center-of-mass energies of until 27 and 100 TeV, respectively. 

On the theoretical side, one of the simplest models reported in the literature is the Two-Higgs-Doublet Model (2HDM) \cite{gunion, cheng}, which offers a good opportunity for the analysis of decay $h\to\tau\mu$. The versions type I and type II of 2HDM are invariant under a $Z_2$ discrete symmetry and due to that some parameters of the scalar potential are complex in general, explicit CP violation can be induced. In particular, the $\lambda_5$ quartic interaction in the Higgs potential can lead to this. In the model type I only one of the doublets gives masses to the fermions \cite{haber}, while in the model type II one doublet is assigned to give mass to the sector up and the other to the sector down, respectively. 
The Two-Higgs-Doublet Model type III (2HDM-III) both doublet scalar fields give masses to the up and down sectors. This general version generate Flavor Changing Neutral Currents (FCNC) in Higgs-fermions Yukawa couplings  and $\mathcal{CP}$ violation ($\mathcal{CPV}$) in the Higgs potential \cite{haber, Fritzsch}. In this paper, we search for the decay $h\to\tau\mu$ in the context of the 2HDM-III. 

The organization of our work is as follows. In section. \ref{SeccionII} we discuss generalities of the 2HDM-III including the Yukawa interaction Lagrangian written in terms of mass eigenstates as well as the diagonalization of the mass matrix. Section \ref{SeccionIII} is devoted to the constraints on the relevant model parameter space whose values will be used in our analysis. The section \ref{SeccionIV} is focused on the analysis of the production cross section of the SM-like Higgs boson via the gluon fusion mechanism, the decay $h\to\tau\mu$ and its possible detection at super hadron colliders, namely: HL-LHC, HE-LHC and the FCC-hh. Finally, conclusions and outlook are presented in section  \ref{SeccionV}.

\section{Two-Higgs Doublet Model type III}
\label{SeccionII}
The 2HDM includes two doublet scalar fields with the same hypercharge, $Y=1$. The classification of the 2HDM types is based on the different ways to introduce Yukawa interactions and scalar potential. In this paper, the theoretical framework adopted is the 2HDM-III, where both doublets are used to induce interactions between fermions and scalars as described in this section. A characteristic of the 2HDM-III is that the fermion mass matrix is a linear combination of two Yukawa matrices, which is diagonalized by a bi-unitarity transformation. However, this bi-unitary transformation do not simultaneously diagonalize the two Yukawa matrices. As a result, FCNC can arise at tree level. 

\subsection{General Higgs potential in the 2HDM-III}
The most general $SU(2)_L\times U(1)_Y$ invariant scalar potential is given by \cite{Gunion:2002zf, Morettietal}:

\begin{eqnarray}\label{potential}
V(\Phi_1, \Phi_2)&=&\mu_1^2(\Phi_1^{\dagger}\Phi_1)+\mu_2^2(\Phi_2^{\dagger}\Phi_2)-\left(\mu_{12}^2(\Phi_1^{\dagger}\Phi_2)+H.c.\right)+\frac{1}{2}\lambda_1(\Phi_1^{\dagger}\Phi_1)^2\\ \nonumber
&+& \frac{1}{2}\lambda_2(\Phi_2^{\dagger}\Phi_2)^2+\lambda_3(\Phi_1^{\dagger}\Phi_1)(\Phi_2^{\dagger}\Phi_2)+\lambda_4(\Phi_1^{\dagger}\Phi_2)(\Phi_2^{\dagger}\Phi_1)\\  \nonumber
&+&\left(\frac{1}{2}\lambda_5(\Phi_1^{\dagger}\Phi_2)^2+\left(\lambda_6(\Phi_1^{\dagger}\Phi_1)+\lambda_7(\Phi_2^{\dagger}\Phi_2)\right)(\Phi_1^{\dagger}\Phi_2)+H.c.\right),
\end{eqnarray} 
where $\mu_{1,\,2}$, $\lambda_{1,\,2,\,3,\,4}$ are real parameters while $\mu_{12}$, $\lambda_{5,\,6,\,7}$ can be complex in general. The doublets are written as $\Phi_{a}^T=\left( \phi_{a}^{+}, \phi_{a}^0\right)$ for $a=1,2$.  After the Spontaneous Symmetry Breaking (SSB) the two Higgs doublets acquire non-zero expectation values. The Vacuum Expectation Values (VEV) are selected as
\begin{equation}
\langle \Phi_a \rangle= \frac{1}{\sqrt{2}}\left(
\begin{array}{c}
0 \\
\upsilon_a \\
\end{array}
\right),\,a=1,\,2;
\label{vev1}
\end{equation}
where $\upsilon_1$ and $\upsilon_2$ satisfy $\upsilon_1^2 + \upsilon_2^2 = \upsilon^2$ for $\upsilon=246$ GeV. Usually, in the 2HDM-I and II the terms proportional to $\lambda_{6,\,7}$ are removed by imposing the $Z_2$ discrete symmetry in which the doublets are transformed as $\Phi_1\to \Phi_1$ and $\Phi_2\to -\Phi_2$. This $Z_2$ discrete symmetry suppresses FCNC in Higgs-fermions Yukawa couplings at tree level. This is the main reason why $Z_2$ discrete symmetry is not introduced in the 2HDM-III.

On the other hand, once the scalar potential (\ref{potential}) is diagonalized, the mass-eigenstates fields are generated. The charged components of $\Phi_a$ lead to a physical charged scalar boson and the pseudo-Goldstone bosons associated with the $W$ gauge fields, these are given as follows:
\begin{eqnarray}
G_W^{\pm}&=&\phi_1^{\pm}\cos\beta+\phi_2^{\pm}\sin\beta,\\
H^{\pm}&=&-\phi_1^{\pm}\sin\beta+\phi_2^{\pm}\cos\beta,
\end{eqnarray}
where the mixing angle $\beta$ is defined through $\tan\beta=\upsilon_2/\upsilon_1(=t_{\beta})$.

The charged scalar boson mass is given by:
\begin{equation}
m_{H^{\pm}}^2=\frac{\mu_{12}^2}{s_{\beta}c_{\beta}}-\frac{1}{2}v^2\left(\lambda_4+\lambda_5+t_{\beta}^{-1}\lambda_6+t_{\beta}\lambda_7\right),
\end{equation}
where we defined $\cos\beta(\sin{\beta})=c_{\beta}(s_{\beta})$.
Meanwhile, the imaginary part of the neutral component of the $\Phi_a$, i.e., $\text{Im}(\Phi^0)$, defines the $\mathcal{CP}$-odd state and the pseudo-Goldstone boson related to the $Z$ gauge boson. The corresponding neutral rotation is given by:
\begin{eqnarray}
G_Z&=&\text{Im}(\phi_1^{0})c_{\beta}+\text{Im}(\phi_2^{0})s_{\beta},\\
A^0&=&-\text{Im}(\phi_1^{0})s_{\beta}+\text{Im}(\phi_2^{0})c_{\beta},
\end{eqnarray}
where the superscript 0 denotes the neutral part of the doubles. The $\mathcal{CP}$-odd scalar boson mass reads as follows:
\begin{equation}
m_{A^0}^2=m_{H^{\pm}}^2+\frac{1}{2}v^2(\lambda_4-\lambda_5).
\end{equation}
On the other side, the real part of the neutral component of the $\Phi_a$, i.e., $\text{Re}(\Phi^0)$, defines the $\mathcal{CP}$-even states, namely: the SM-like Higgs boson $h$ and a heavy scalar boson $H$. 

The physical $\mathcal{CP}$-even states are written as:
\begin{eqnarray}
H&=&\text{Re}(\phi_1^0)c_{\alpha}+\text{Re}(\phi_2^0)s_{\alpha},\\
h&=&-\text{Re}(\phi_1^0)s_{\alpha}+\text{Re}(\phi_2^0)c_{\alpha},
\end{eqnarray}
with
\begin{equation}
\tan2\alpha=\frac{2m_{12}}{m_{11}-m_{22}},
\end{equation}
where $m_{11}$, $m_{12}$, $m_{22}$ are elements of the real part of the mass matrix $\bf{M}$,
\begin{equation}
\text{Re}(\bf{M})=\left(\begin{array}{cc}
m_{11} & m_{12}\\
m_{12} & m_{22}
\end{array}\right),
\end{equation}
with:
\begin{eqnarray}
m_{11}&=&m_A^2 s_{\beta}^2+\upsilon^2\left(\lambda_1 c_{\beta}^2+\lambda_5 s_{\beta}^2+2\lambda_6 c_{\beta}s_{\beta}\right),\\
m_{12}&=&-m_A^2 c_{\beta} s_{\beta}+\upsilon^2\left[(\lambda_3+\lambda_4) c_{\beta} s_{\beta}+\lambda_6 c_{\beta}^2+\lambda_7 s_{\beta}^2\right],\\
m_{22}&=&m_A^2 c_{\beta}^2+\upsilon^2\left(\lambda_2 s_{\beta}^2+\lambda_5 c_{\beta}^2+2\lambda_7 c_{\beta}s_{\beta}\right).
\end{eqnarray}

Finally, the neutral $\mathcal{CP}$-even scalar masses are written as follows:
\begin{equation}
m_{H,\,h}^2=\frac{1}{2}\left(m_{11}+m_{22}\pm \sqrt{(m_{11}-m_{22})^2+4m_{12}^2}\right).
\end{equation}

\subsection{Yukawa Lagrangian of the THDM-III}
In the most general case both doublets can participate in the interactions with the fermion fields. The Yukawa Lagrangian is written as
\begin{eqnarray}\label{Yukawa}
\mathcal{L}_{Y} & = & Y_{1}^{u}\bar{Q}_{L}^{'}\tilde{\Phi}_{1}u_{R}^{'}+Y_{2}^{u}\bar{Q}_{L}^{'}\tilde{\Phi}_{2}u_{R}^{'}+Y_{1}^{d}\bar{Q}_{L}^{'}\Phi_{1}d_{R}^{'}\nonumber \\
 & + & Y_{2}^{d}\bar{Q}_{L}^{'}\Phi_{2}d_{R}^{'}+Y_{1}^{\ell}\bar{L}_{L}^{'}\Phi_{1}\ell_{R}^{'}+Y_{2}^{\ell}\bar{L}_{L}^{'}\Phi_{2}\ell_{R}^{'}+H.c.,
\end{eqnarray}
with
\begin{eqnarray}\label{terminosLagrangiano}
Q_{L}^{'} & = & \left(\begin{array}{c}
u_{L}^{'}\\ \nonumber
d_{L}^{'}
\end{array}\right),\;L_L^{'}=\left(\begin{array}{c}
\nu_{L}^{'}\\
e_{L}^{'}
\end{array}\right),\\
\Phi_{1} & = & \left(\begin{array}{c}
\phi_{1}^{+}\\
\phi_{1}^{'}
\end{array}\right),\;\Phi_{2}=\left(\begin{array}{c}
\phi_{2}^{+}\\
\phi_{2}^{'}
\end{array}\right),\\
\tilde{\Phi}_{j} & = & i\sigma_{2}\Phi_{j}^{*}.\nonumber
\end{eqnarray}
The apostrophe superscript in fermion fields stands for the interaction basis. The left-handed doublets and right-handed singlets are denoted with the subscripts $L$ and $R$, respectively. $Y_{i}^{f}$ ($i=1,\,2$; $f=u,\,d,\,\ell$) are the $3\times 3$ Yukawa matrices.

Introducing the expressions \eqref{terminosLagrangiano} in \eqref{Yukawa} and after the SSB,  the neutral Yukawa Lagrangian is given by:

\begin{eqnarray}\label{Yukawa2}
{\cal{L}}_Y^{0} &=& \bar{u}^{'}\frac{1}{\sqrt{2}}\left(\upsilon_1 Y_1^u+\upsilon_2 Y_2^u\right)u^{'}+\bar{d}^{'}\frac{1}{\sqrt{2}}\left(\upsilon_1 Y_1^d+\upsilon_2 Y_2^d\right)d^{'} \\ \nonumber
&+&\bar{u}^{'}\left[\frac{1}{\sqrt{2}}\left(Y_1^u c_{\alpha}+Y_2^u s_{\alpha}\right)H 
+\frac{1}{\sqrt{2}}\left(-Y_1^u s_{\alpha}+Y_2^u c_{\alpha}\right)h 
+ i\frac{1}{\sqrt{2}}\left(Y_1^u s_{\beta}-Y_2^u c_{\beta}\right)\gamma^5 A \right] u^{'}\\ \nonumber
&+&\bar{d}^{'}\left[\frac{1}{\sqrt{2}}\left(Y_1^d c_{\alpha}+Y_2^d s_{\alpha}\right)H 
+\frac{1}{\sqrt{2}}\left(-Y_1^d s_{\alpha}+Y_2^d c_{\alpha}\right)h 
+ i \frac{1}{\sqrt{2}}\left(-Y_1^d s_{\beta}+Y_2^d c_{\beta}\right)\gamma^5 A\right] d^{'}\\ \nonumber
\end{eqnarray}

The first two terms are associated with the masses of the fermion particles, as we will see below; while the rest define the couplings of the scalar bosons with fermion pairs. The corresponding charged lepton part is obtained by replacing $d\to\ell$.

\subsubsection{Diagonalization of the fermion mass matrices}
The first two terms of eq. \eqref{Yukawa2} are associated to the fermion mass matrices:
\begin{equation}
 M_f=\frac{1}{\sqrt{2}}\left( v_{1} Y_{1}^{f}+v_{2}Y_{2}^{f}\right),\;f=u,\,d,\,\ell. 
\end{equation}
 
We assume that mass matrices have a structure of four zero textures \cite{Fritzsch123, Branco, Cruz:2019vuo, LorPapRos, Arroyo}, namely: 
\begin{equation}\label{4zero_MASS_matrices}
M_f=\left(\begin{array}{ccc}0 & D_f & 0 \\D_f & C_f & B_f \\0 & B_f & A_f\end{array}\right),\;
\end{equation}
The elements of a real matrix of the type \eqref{4zero_MASS_matrices}
are related to the eigenvalues $m_i$, ($i=1,\,2,\,3$) \cite{Branco}, through the following invariants:
\begin{eqnarray}\label{Invariantes}
det\left(M\right) & = & -D^{2}A=m_{1}m_{2}m_{3},\nonumber \\
Tr\left(M\right) & = & C+A=m_{1}+m_{2}+m_{3},\\
\lambda\left(M\right) & = & CA-D^{2}-B=m_{1}m_{2}+m_{1}m_{3}+m_{2}m_{3},\nonumber 
\end{eqnarray} 
where we have omitted the subscript $f$ to not overload the notation.
From eqs. \eqref{Invariantes} we find a relation between the components of the mass matrix of four zero textures and the eigenvalues $m_i$ ($i=1,\,2,\,3$), namely:
\begin{eqnarray}\label{ElementosMatrizMasa}
A & = & m_{3}-m_2,\nonumber \\
B & = & m_{3}\sqrt{\frac{r_{2}(r_{2}+r_{1}-1)(r_{2}+r_{2}-1)}{1-r_{2}},}\\
C & = & m_{3}(r_{2}+r_{1}+r_{2}),\nonumber \\
D & = & \sqrt{\frac{m_{1}m_{2}}{1-r_{2}}},\nonumber 
\end{eqnarray}
with $r_i=m_i/m_3$. 

On the other side, without losing generality, a hierarchy between the eigenvalues $m_i$ such that |$m_1$|<|$m_2$|<|$m_3$| and 0<$m_2$<A<$m_3$, is assumed. Under these considerations, the mass matrix can be diagonalized by the bi-unitary transformation $\bar{M}_f=V_{fL}^{\dagger}M_f V_{fR}=\textrm{Diag}\left\{m_{f_1},\,m_{f_2},\,m_{f_3}\right\}$. The fact that $M_f$ is hermitian, implies that $V_{fL}=V_{fR}\equiv V_f$ which is given by  $V_f=\mathcal{O}_{f}P_f$, with $P_f=\text{Diag}\{e^{i\alpha_f},\,e^{i\beta_f},\,1\}$ and  
\begin{equation}
\mathcal{O}_{f}=\left(\begin{array}{ccc}
\sqrt{\frac{m_{f_2}m_{f_3}(A-m_{f_1})}{A(m_{f_2}-m_{f_1})(m_{f_3}-m_{f_1})}} & \sqrt{\frac{m_{f_1}m_{f_3}(m_{f_2}-A)}{A(m_{f_2}-m_{f_1})(m_{f_3}-m_{f_2})}} & \sqrt{\frac{m_{f_1}m_{f_3}(A-m_{f_3})}{A(m_{f_3}-m_{f_1})(m_{f_3}-m_{f_2})}}\\
-\sqrt{\frac{m_{f_1}(m_{f_1}-A)}{(m_{f_2}-m_{f_1})(m_{f_3}-m_{f_1})}} & \sqrt{\frac{m_{f_2}(A-m_{f_2})}{(m_{f_2}-m_{f_1})(m_{f_3}-m_{f_2})}} & \sqrt{\frac{m_{f_3}(m_{f_2}-A)}{(m_{f_2}-m_{f_1})(m_{f_3}-m_{f_2})}}\\
\sqrt{\frac{m_{f_1}(A-m_{f_2})(A-m_{f_3})}{A(m_{f_2}-m_{f_1})(m_{f_3}-m_{f_1})}} & -\sqrt{\frac{m_{f_2}(A-m_{f_1})(m_{f_3}-A)}{A(m_{f_2}-m_{f_1})(m_{f_3}-m_{f_2})}} & \sqrt{\frac{m_{f_3}(A-m_{f_1})(A-m_{f_2})}{A(m_{f_3}-m_{f_1})(m_{f_3}-m_{f_2})}}
\end{array}\right),
\end{equation}
where we identify to $m_{f_i}$ $(i=1,\,2,\,3)$ as the physical fermion masses.
 A remarkable fact is that $V_{f}$ must reproduces the observed CKM matrix elements ($V_{\text{CKM}}$), which is achieved as $V_{\text{CKM}}=V_{u}^{\dagger}V_{d}$. In Ref. \cite{Branco} and in a previous research by one of us \cite{Arroyo} a numerical analysis was presented, in which the $V_{\text{CKM}}$ matrix is reproduced satisfactorily. It is worth mentioning that the CP phase can be identified through the matrix $P_f=\text{Diag}\{e^{i\alpha_f},\,e^{i\beta_f},\,1\}$.

Once the bi-unitary transformation is applied, the fermion mass matrix is transformed as 
\begin{equation}
\bar{M}_f =\frac{v_{1}}{\sqrt{2}}\tilde{Y}_{1}+\frac{v_{2}}{\sqrt{2}}\tilde{Y}_{2},\;  \tilde{Y}_{1,2}=V_f^{\dagger}Y_{1,2}V_f. \label{mass}
\end{equation}
Unitary matrices only diagonalize to the fermion mass matrices $M_f$, leaving Yukawa matrices, in general, as non-diagonal. Then, FCNC are induced at tree level. 
\subsubsection{Flavor-changing neutral scalar interactions}
The eq. \eqref{mass} not only defines the mass matrix but also provide relations between the Yukawa matrices. In order to obtain the interactions in terms of only one Yukawa matrix, the eq. \eqref{mass} can be written in two possible forms:
\begin{eqnarray}
\tilde{Y}^f_1 &=&\frac{\sqrt{2}}{v_1}\bar{M}_{f}-\tan\beta \tilde{Y}^f_{2} \\
\tilde{Y}^f_2&=&\frac{\sqrt{2}}{v_2}\bar{M}_{f}-\cot\beta \tilde{Y}^f_{1}.
\end{eqnarray}
On the other side, the Yukawa Lagrangian \eqref{Yukawa2} after being expanded in terms of mass eigenstates, which is achieved with the transformations $f_L=V_{fL}^{\dagger}f^{'}$, $f_R=V_{fR}^{\dagger}f^{'}$, can be written in different versions \cite{Gaitan:2017tka}, however, we choose to write the Yukawa interactions as a function of $\tilde{Y}_2$. From now on, in order to simplify the notation, the subscript 2 in the Yukawa couplings will be omitted.

The interactions between fermions and the neutral scalar bosons are explicitly written as
\begin{eqnarray}\label{Full_Neutral_YukawaLagrangian}
{\cal{L}}_Y & = &
\frac{g}{2}\left(\frac{m_{\ell_i}}{m_W}\right)\bar{\ell}_{i}
\left[-\frac{\sin\alpha}{\cos\beta} \delta_{ij}+  
\frac{\sqrt{2}\cos(\alpha - \beta)}{g\cos\beta}\left(\frac{m_W}{m_{\ell_i}}\right) \tilde{Y}_{ij}^{\ell}\right]\ell_{j}h \nonumber \\
&+& 
\frac{g}{2}\left(\frac{m_{\ell_i}}{m_W}\right)\bar{\ell}_{i}
\left[\frac{ \cos\alpha}{\cos\beta}\delta_{ij}+
\frac{\sqrt{2}\sin(\alpha - \beta)}{g\cos\beta}\left(\frac{m_W}{m_{\ell_i}}\right) \tilde{Y}_{ij}^{\ell}\right]\ell_{j}H \nonumber \\
& +&i
\frac{g}{2}\left(\frac{m_{\ell_i}}{m_W}\right)\bar{\ell}_{i} \left[-\tan\beta \delta_{ij}+  \frac{\sqrt{2}}{ g\cos\beta}\left(\frac{m_W}{m_{\ell_i}}\right)\tilde{Y}_{ij}^{\ell}\right]
\gamma^{5}\ell_{j} A \nonumber \\
&+&
\frac{g}{2}\left(\frac{m_{d_i}}{m_W}\right)\bar{d}_{i}
\left[-\frac{\sin\alpha}{\cos\beta} \delta_{ij}+  
\frac{\sqrt{2}\cos(\alpha - \beta)}{g\cos\beta}\left(\frac{m_W}{m_{d_i}}\right) \tilde{Y}_{ij}^{d}\right]d_{j}h \nonumber \\
&+& 
\frac{g}{2}\left(\frac{m_{d_i}}{m_W}\right)\bar{d}_{i}
\left[\frac{ \cos\alpha}{\cos\beta}\delta_{ij}+
\frac{\sqrt{2}\sin(\alpha - \beta)}{g\cos\beta}\left(\frac{m_W}{m_{d_i}}\right) \tilde{Y}_{ij}^{d}\right]d_{j}H \nonumber \\
& +&i
\frac{g}{2}\left(\frac{m_{d_i}}{m_W}\right)\bar{d}_{i} \left[-\tan\beta \delta_{ij}+  \frac{\sqrt{2}}{ g\cos\beta}\left(\frac{m_W}{m_{d_i}}\right)\tilde{Y}_{ij}^{d}\right]
\gamma^{5}d_{j} A \nonumber \\ 
 & + & \frac{g}{2}\left(\frac{m_{u}}{m_{W}}\right)\bar{u}_{i}\left[\frac{\sin\alpha}{\sin\beta}\delta_{ij}+\frac{\sqrt{2}\sin(\alpha-\beta)}{g\sin\beta}\left(\frac{m_{W}}{m_{u}}\right)\tilde{Y}^{u}_{ij}\right]u_{j}H \nonumber \\ 
 & + & \frac{g}{2}\left(\frac{m_{u}}{m_{W}}\right)\bar{u}_{i}\left[\frac{\cos\alpha}{\sin\beta}\delta_{ij}-\frac{\sqrt{2}\cos(\alpha-\beta)}{g\sin\beta}\left(\frac{m_{W}}{m_{u}}\right)\tilde{Y}^{u}_{ij}\right]u_{j}h\nonumber \\
 &+&i\frac{g}{2}\left(\frac{m_{u}}{m_{W}}\right)\bar{u}_{i}\left[-\cot\beta\delta_{ij}+\frac{\sqrt{2}}{g\sin\beta}\left(\frac{m_{W}}{m_{u}}\right)\tilde{Y}^{u}_{ij}\right]\gamma^{5}u_{j}A,
\end{eqnarray}
where $i$ and $j$ stand for the fermion flavors, in general $i\neq j$.
The first term in eq. \eqref{Full_Neutral_YukawaLagrangian} between brackets corresponds to the contribution of the THDM-II over the SM result, while the term proportional to $\tilde{Y}_{ij}^f$ is the new contribution from the THDM-III. Finally, from eq. \eqref{mass}, the rotated Yukawa matrices $\tilde{Y}_{ij}^f$ are given by:
\begin{equation}
\tilde{Y}_{ij}= \frac{\sqrt{m_i m_j}}{\upsilon}\chi_{ij},
\label{ansatz}
\end{equation}
i. e., the Cheng-Sher ansatz \cite{Cheng:1987rs} times the factor $\chi_{ij}$, which is expected to be of the order of one.

\section{Model parameter space}
\label{SeccionIII}
\begin{figure}[!ht]
\center{\includegraphics[scale=0.6]{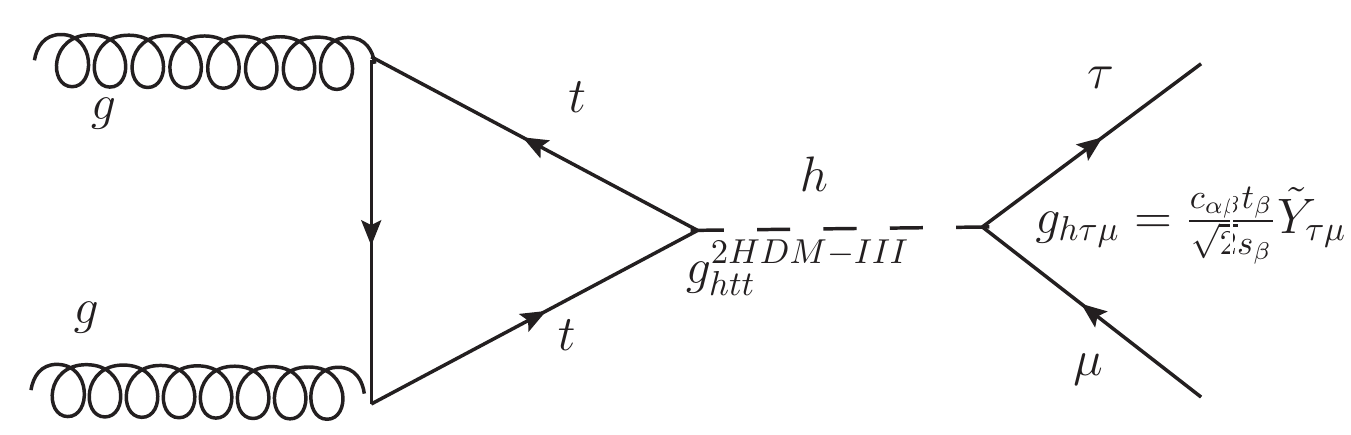}}
 \caption{Feynman diagram of the Higgs boson production via the gluon fusion mechanism with its subsequent decay into $\tau\mu$ pair. The $g_{htt}^{\text{2HDM-III}}$ coupling can be consulted in eq. \eqref{Full_Neutral_YukawaLagrangian}. \label{FeynmanDiagram}   }
\end{figure}

 In order to evaluate the branching ratio of the $h\to\tau\mu$ decay and the production cross section of the SM-like Higgs boson by the gluon fusion mechanism, we need to analyze the 2HDM-III free model parameter space. The most relevant 2HDM-III parameters involved in this work are the $\cos(\alpha-\beta)=c_{\alpha\beta}$ and $\tan\beta=t_{\beta}$ because $g_{h\tau\mu}$ and $g_{htt}$ couplings are proportional to them. Figure \ref{FeynmanDiagram} illustrates this. 
 
To constrain the above mentioned parameters, we consider the LHC Higgs boson data, the decay $B_s^0\to\mu^-\mu^+$, the tau lepton  decays $\tau\to\bar{\ell}_i\bar{\ell}_j\ell_j$ and $\tau\to\ell_i\gamma$ as well as the experimental constraint on the $h\to\tau\mu$ and the muon anomalous magnetic dipole moment $\delta a_{\mu}$. Direct searches for additional heavy neutral $\mathcal{CP}$-even and $\mathcal{CP}$-odd scalars through $gb\to\phi\to\tau\tau$ \cite{ATLASCONF2017, Sirunyan:2018zut}, with $\phi=\,H,\,A$ are also used in order to constrain their masses, we denote them as $m_H$, $m_A$. Finally, the charged scalar boson mass $m_{H^{\pm}}$ is constrained with the upper limit on $\sigma(pp\to tbH^{\pm})\times \mathcal{BR}(H^{\pm}\to\tau^{\pm}\nu)$ \cite{Aaboud:2018gjj} and the decay $b\to s\gamma$ \cite{Chetyrkin:1996vx, Adel:1993ah, Ali:1995bi, Greub:1996tg, Ciuchini:1997xe, Misiak:2017bgg}.

\subsection{Constraint on 
	\texorpdfstring{$c_{\alpha\beta}$}{cab}
	and 
	\texorpdfstring{$t_{\beta}$}{tb}
}

 In order to have values of $c_{\alpha\beta}$ in accordance with current experimental results, we use the coupling modifiers $\kappa$-factors reported by ATLAS and CMS collaborations \cite{ATLAS:2018doi, Sirunyan:2018koj}. They are defined as following:
\begin{equation}
\kappa_{pp}^2=\frac{\sigma(pp\to h^{\text{2HDM-III}})}{\sigma(pp\to h^{\text{SM}})}\;\text{or}\;\kappa_{x\bar{x}}^2=\frac{\Gamma(h^{\text{2HDM-III}}\to x\bar{x})}{\Gamma({h^\text{SM}}\to x\bar{x})}.
\end{equation}
where $\Gamma(H_i\to x\bar{x})$ is the decay width of $H_i$ into $x\bar{x}=b\bar{b},\,\tau^-\tau^+,\,ZZ,\,WW,\,\gamma\gamma$ and $gg$; with $H_i=h^{\text{2HDM-III}}$ and $h^{\text{SM}}$. Here $h^{\text{2HDM-III}}$ is the SM-like Higgs boson coming from 2HDM-III and $h^{\text{SM}}$ is the SM Higgs boson; $\sigma(pp\to H_i)$ is the Higgs boson production cross section via proton-proton collisions. In addition, we also consider the current experimental limits on the tau decays $\tau\to\mu\gamma$, $\tau\to\bar{\ell}_i\bar{\ell}_j\ell_j$, $\delta a_{\mu}$, $B_s^0\to \mu^-\mu^+$ \cite{PDG} and the direct upper bound on the branching ratio of the Higgs boson into $\tau\mu$ pair \cite{Sirunyan:2017xzt, ATLAS:2019icc}. All the necessary formulas to perform our analysis of the model parameter space are presented in Appendix \ref{FormulaDecays}.

 In figure \ref{tb_cab} we present the $c_{\alpha\beta}-t_{\beta}$ planes in which the shadowed areas represent the allowed regions by:
 \begin{itemize}
 \item[\ref{2a}] The decay $B_s^0\to\mu^+\mu^-$,
 \item[\ref{2b}] Coupling modifiers $\kappa_X$,
 \item[\ref{2c}] Lepton Flavor Violating Processes: $\tau\to\mu\gamma$, $\tau\to\bar{\ell}_i\bar{\ell}_j\ell_j$, $\delta a_{\mu}$ and $h\to\tau\mu$,
 \item [\ref{2d}] Intersection of all individual allowed regions in which we display both the most up-to-date results reported by LHC and the expected results at the HL-LHC and HE-LHC for Higgs boson data \cite{Cepeda:2019klc} and for the decay $B_s^0\to\mu^+\mu^-$ \cite{ATLAS_REPORT_Bmumu-HLlhc}.
 \end{itemize}  
 We find strong restrictions for the 2HDM-III parameter space on the $c_{\alpha\beta}-t_{\beta}$ plane. We observe that $c_{\alpha\beta}\approx 0.05$ admits a value of $t_{\beta}\approx 8$ for all cases, while $c_{\alpha\beta}=0$ allows $t_{\beta}\approx 12,\,11,\,10$ for the LHC, HL-LHC and HE-LHC, respectively. The graphics were generated with the package $\texttt{SpaceMath}$  \cite{SpaceMath}.
An important point is the fact that the 2HDM-III is able to accommodate the current discrepancy between the theoretical SM prediction and the experimental measurement of the muon anomalous magnetic dipole moment $\delta a_{\mu}$. However, from figure \ref{tb_cab}, we note that the allowed region by $\delta a_{\mu}$ is out of the intersection of the additional observables. This happens by choosing the parameters shown in table \ref{ParValues}. We find that $\delta a_{\mu}$ is sensitive to $\chi_{\tau\mu}$ which is set to the unit in order to obtain the best fit of the model parameter space. Under this choice, $\delta a_{\mu}$ is explained with high values of $t_{\beta}$.

\begin{figure}[!ht]
\centering
\subfigure[]{{\includegraphics[scale=0.148]{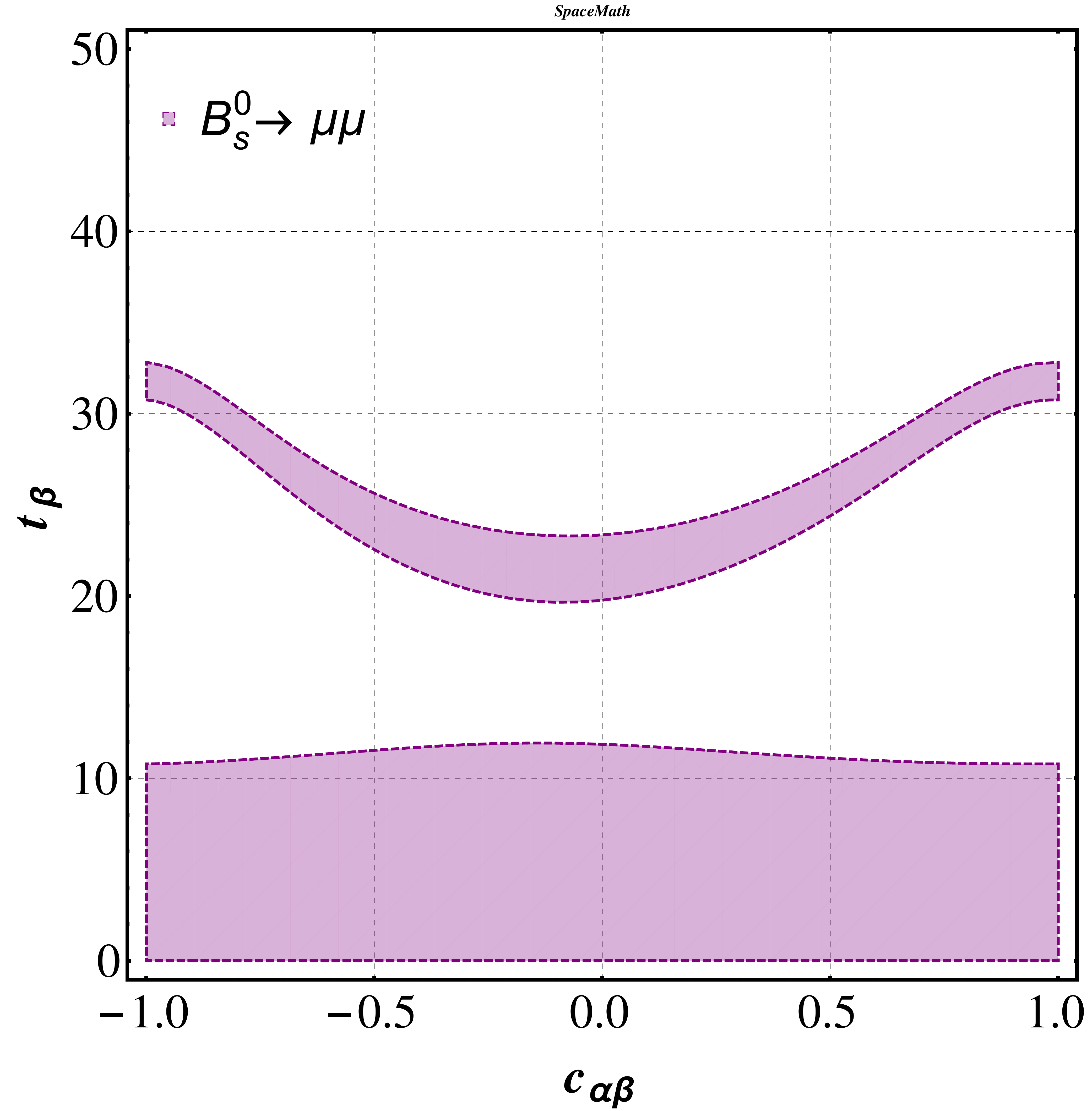}}\label{2a}}
\subfigure[]{{\includegraphics[scale=0.17]{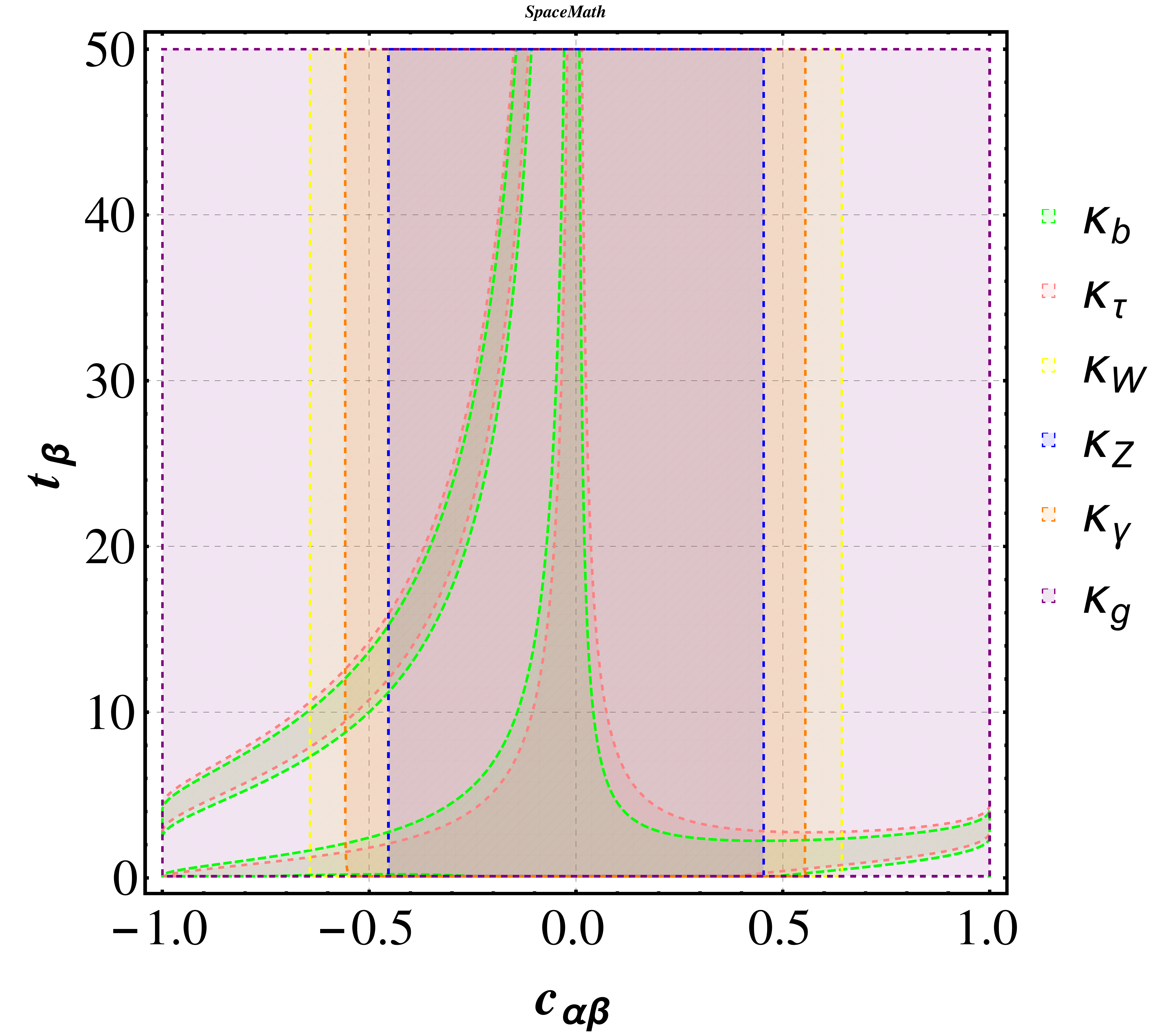}}\label{2b}}
\subfigure[]{{\includegraphics[scale=0.185]{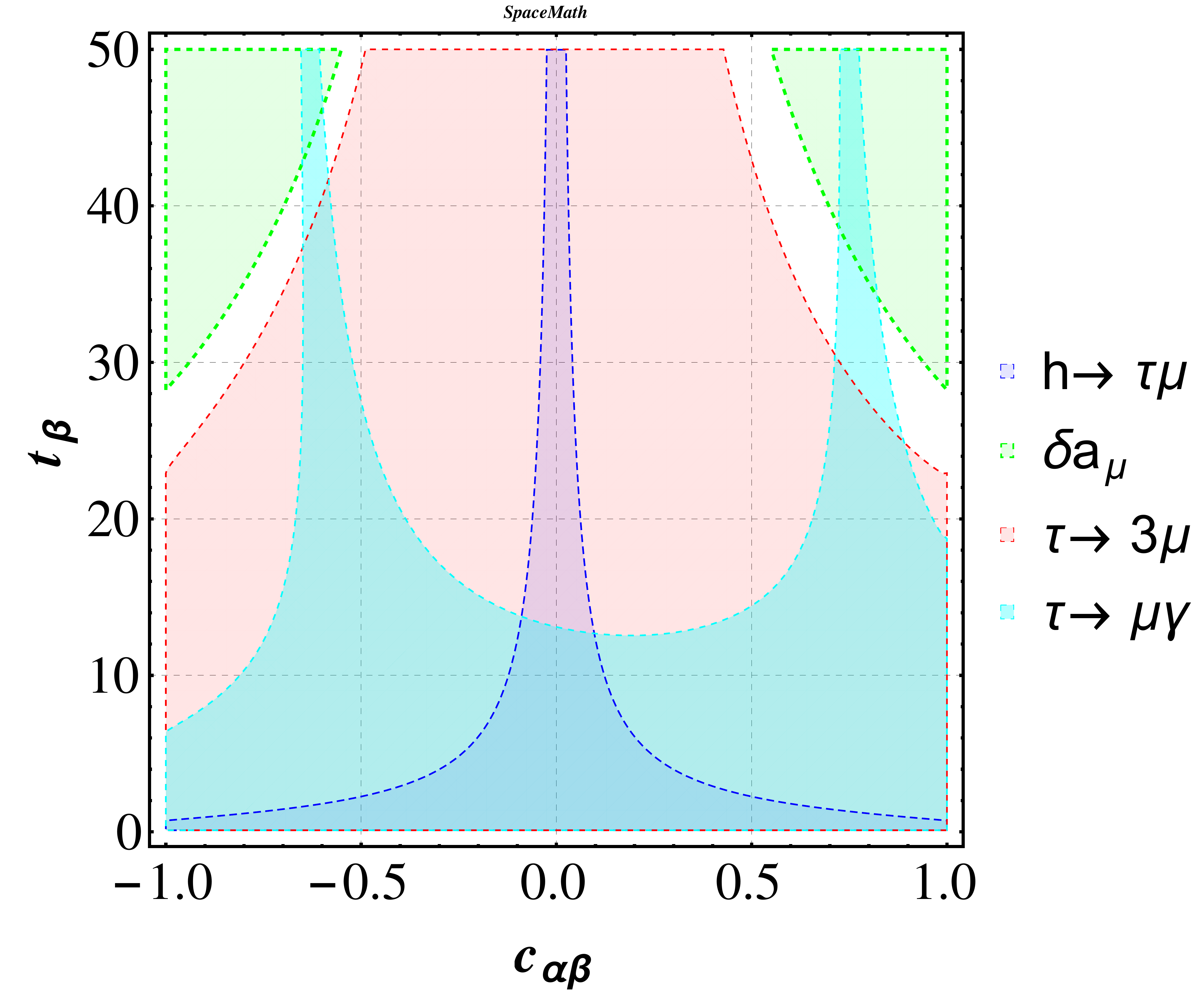}}\label{2c}}
\subfigure[]{{\includegraphics[scale=0.2]{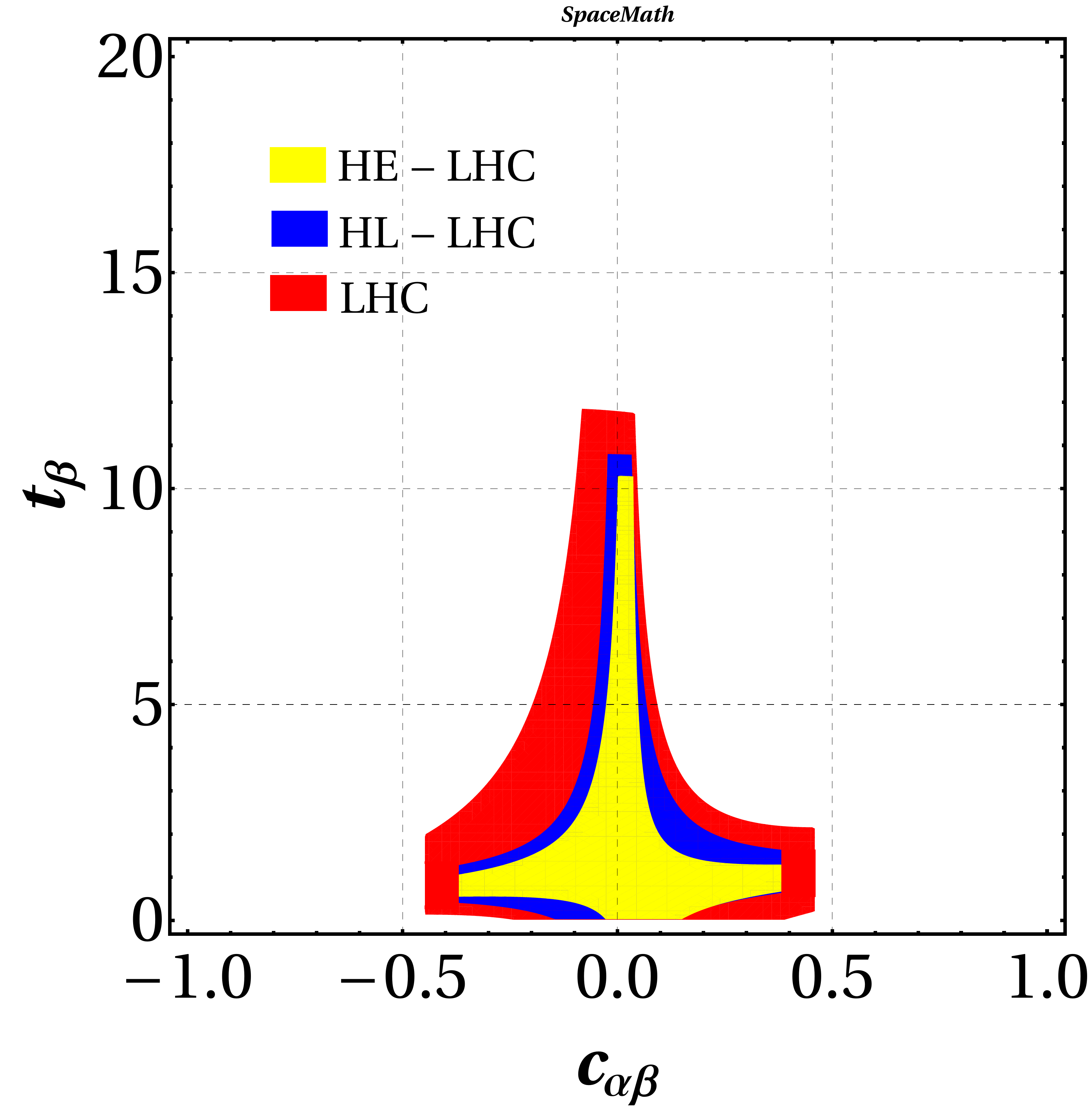}}\label{2d}}
 \caption{The shadowed areas represent the allowed regions in the plane $c_{\alpha\beta}$-$t_{\beta}$: (a) $B_s^0\to\mu^+\mu^-$, (b) Coupling modifiers $\kappa$-factors, (c) LFV processes and (d) Intersection of all allowed regions in which we show the cases for the LHC, HL-LHC and HE-LHC.   \label{tb_cab}  }
\end{figure}

\subsection{Constraint on 
	\texorpdfstring{$m_H$}{mH}, 
	\texorpdfstring{$m_A$}{mA} 
	and 
	\texorpdfstring{$m_{H^{\pm}}$}{mH+-}
}
\subsubsection{
	\texorpdfstring{$m_H$}{mH} 
	and 
	\texorpdfstring{$m_A$}{mA}
}
\begin{figure}[!ht]
\center{\includegraphics[scale=0.6]{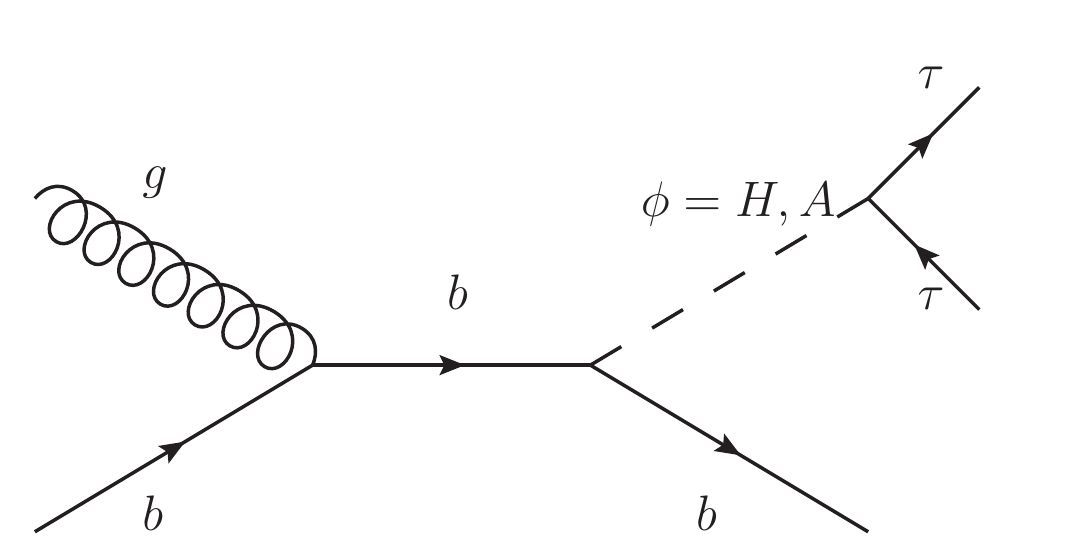}}
 \caption{Feynman diagram of the production of $\phi$ in association with a bottom quark at LHC, with a subsequent decay into $\tau\tau$ pair.  \label{FeynmanDiagram2}   }
\end{figure}

The ATLAS and CMS collaborations presented results of a search for additional neutral Higgs bosons in the ditau decay channel \cite{ATLASCONF2017, Sirunyan:2018zut}. The former of them searched through the process $gb\to\phi\to\tau\tau$, with $\phi=A,\,H$;  figure \ref{FeynmanDiagram2} shows the Feynman diagram of this reaction.  However, no evidence of any additional Higgs boson was observed. Nevertheless, upper limits on the production cross section $\sigma(gb\to\phi)$ times branching ratio $\mathcal{BR}(\phi\to\tau\tau)$ were imposed.
 In this work we focus on the particular case of the search carried out by the ATLAS collaboration. 

\begin{figure}[!ht]
\centering
\subfigure[ ]{\includegraphics[scale=0.295,angle=270]{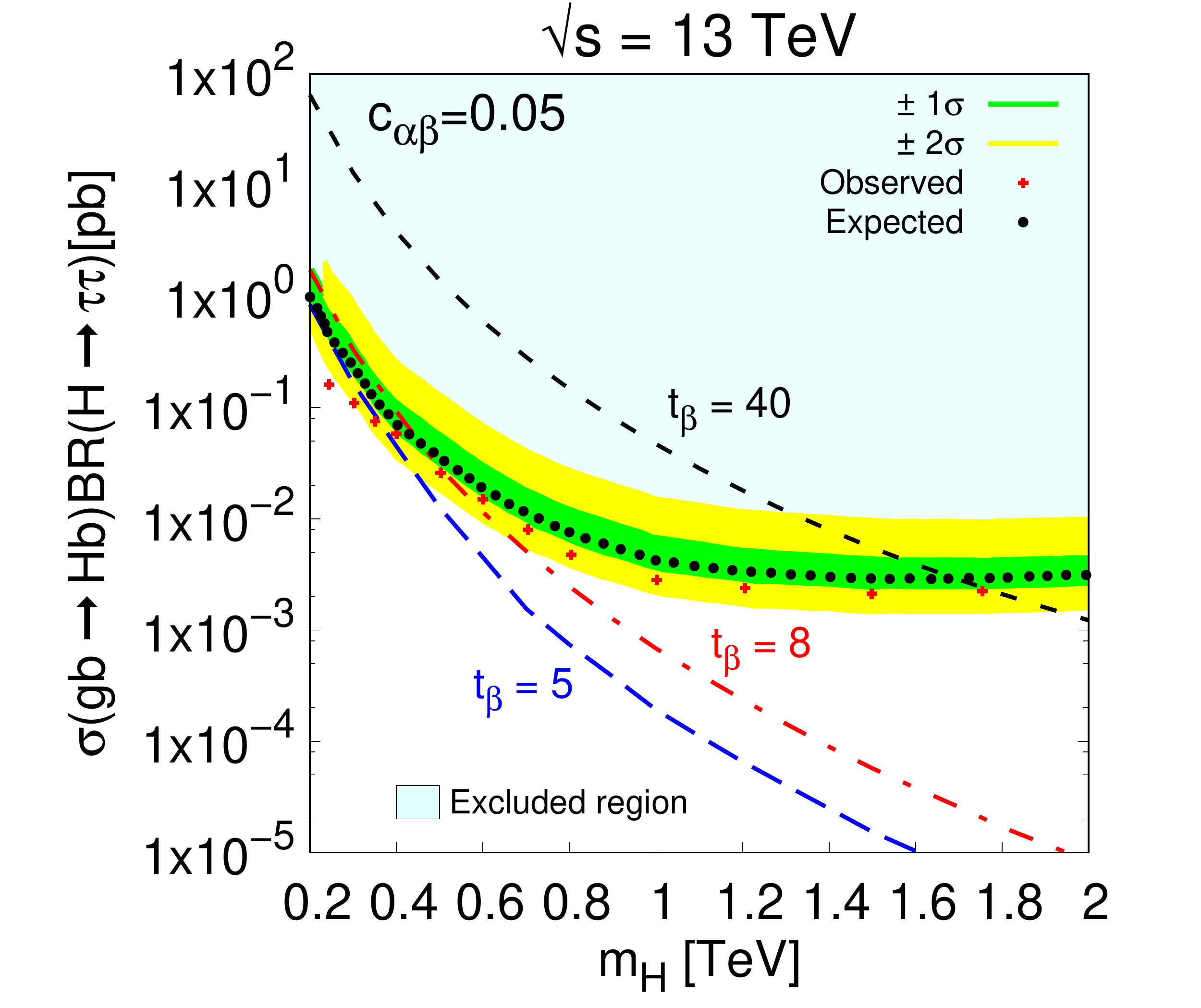}\label{XS_H_timesBRHtautau}}
\subfigure[ ]{\includegraphics[scale=0.295,angle=270]{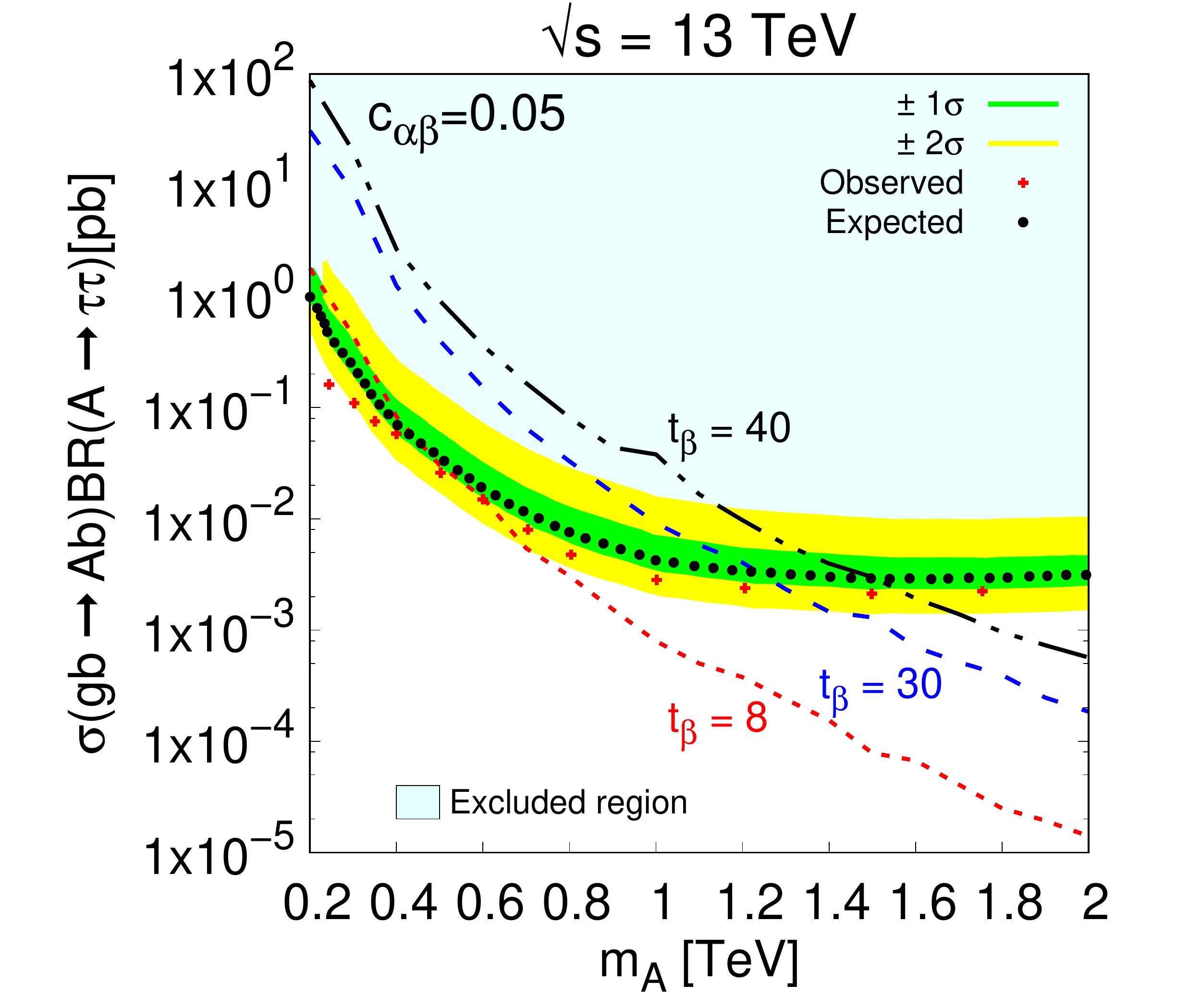}\label{XS_A_timesBRHtautau}}
 \caption{The observed and expected at 95$\%$ CL upper limits on the production cross section times ditau branching ratio for a scalar boson produced via $b$-associated production as a function of (a) the $\mathcal{CP}$-even mass for $t_{\beta}=$5, 8, 40 and (b) the $\mathcal{CP}$-odd mass for $t_{\beta}=$ 8, 30, 40. We take $c_{\alpha\beta}=0.05$.    }
\end{figure}

In figure \ref{XS_H_timesBRHtautau}, we present the $\sigma(gb\to Hb)\times\mathcal{BR}(H\to\tau\tau)$ as a function of $m_H$ for illustrative values of $t_{\beta}=5,\,8,\,40$ and $c_{\alpha\beta}=0.05$. Figure \ref{XS_A_timesBRHtautau} shows the same but as a function of $m_A$ and values for $t_{\beta}=8,\,30,\,40$. In both plots, the black points and red crosses represent the expected and observed values at 95$\%$ CL upper limits, respectively; while the green (yellow) band indicates the interval at $\pm 1 \sigma$ ($\pm 2 \sigma$) with respect to the expected value. We implement the Feynman rules in $\texttt{CalcHEP}$ \cite{CalcHEP} in order to evaluate $\sigma(gb\to \phi b)\times\mathcal{BR}(\phi\to\tau\tau)$. 

From figure \ref{XS_H_timesBRHtautau} we note that $m_H\lesssim 690$ GeV ($m_H\lesssim 510$ GeV) are excluded at 2$\sigma$ (1$\sigma$) for $t_{\beta}= 8$, while for $t_{\beta}\lesssim 4$ the upper limit on $\sigma(gb\to \phi b)\times\mathcal{BR}(\phi\to\tau\tau)$ is easily accomplished. Although $t_{\beta}=$40 is discarded, as shown in figure \ref{tb_cab}, we include it to have an overview of the behavior of the model. On the other side, from figure \ref{XS_A_timesBRHtautau}, we observe that $m_A\lesssim 710$ GeV ($m_H\lesssim$ 610 GeV) are excluded at 2$\sigma$ (1$\sigma$) for $t_{\beta}= 8$. 

\subsubsection{Constraint on the charged scalar mass 
	\texorpdfstring{$m_{H^\pm}$}{mH+-}
}

The discovery of a charged scalar $H^{\pm}$ would constitute unambiguous evidence of new physics.
Direct constraints can be obtained from collider searches for the production and decay of on-shell charged Higgs bosons. These limits are very robust and model-independent if the basic assumptions on the production and decay modes are satisfied \cite{Khachatryan:2015qxa, Aad:2013hla, Khachatryan:2015uua, CMS:2016qoa}. More recently the ATLAS collaboration reported a study on the charged Higgs boson produced either in top-quark decays or in association with a top quark. Subsequently the charged Higgs boson decays via $H^{\pm}\to\tau^{\pm}\nu_{\tau}$ with a center-of-mass energy of 13 TeV \cite{Aaboud:2018gjj}. We analyze this process through the $\texttt{CalcHEP}$ package, however, we find that this process is not a good way to impose a stringent bound on $m_{H^{\pm}}$.

 Conversely, the decay $b\to s\gamma$ imposes stringent limits on $m_{H^\pm}$ because a new ingredient with respect to the SM contribution \cite{Chetyrkin:1996vx, Adel:1993ah, Ali:1995bi, Greub:1996tg} is the presence of the charged scalar boson coming from 2HDM-III which gives contributions to the Wilson coefficients of the effective theory as is shown in the Refs. \cite{Ciuchini:1997xe, Misiak:2017bgg}. 

 In figure 5 we show $R_{quark}$ at NLO in QCD as a function of the charged scalar boson mass for $t_{\beta} = 2, 5, 10$, where $R_{quark}$ is defined as following:
\begin{equation}
R_{quark}=\frac{\Gamma(b\to X_s\gamma)}{\Gamma(b\to X_ce\nu_e)}.
\end{equation} 
 
  We observe that for $t_{\beta}=2$, the charged scalar boson mass $100\,\text{GeV}\lesssim m_{H^{\pm}}$ ($700\,\text{GeV}\lesssim m_{H^{\pm}}$) is excluded, at $2\sigma$ ($1\sigma$); while $t_{\beta}=10$ imposes a more restrictive lower bound $1.6\,\text{TeV}\lesssim m_{H^{\pm}}$ ($3.2\,\text{TeV}\lesssim m_{H^{\pm}}$) at $2\sigma$ ($1\sigma$). 
\begin{figure}[!ht]
\center{\includegraphics[scale=0.4,angle=270]{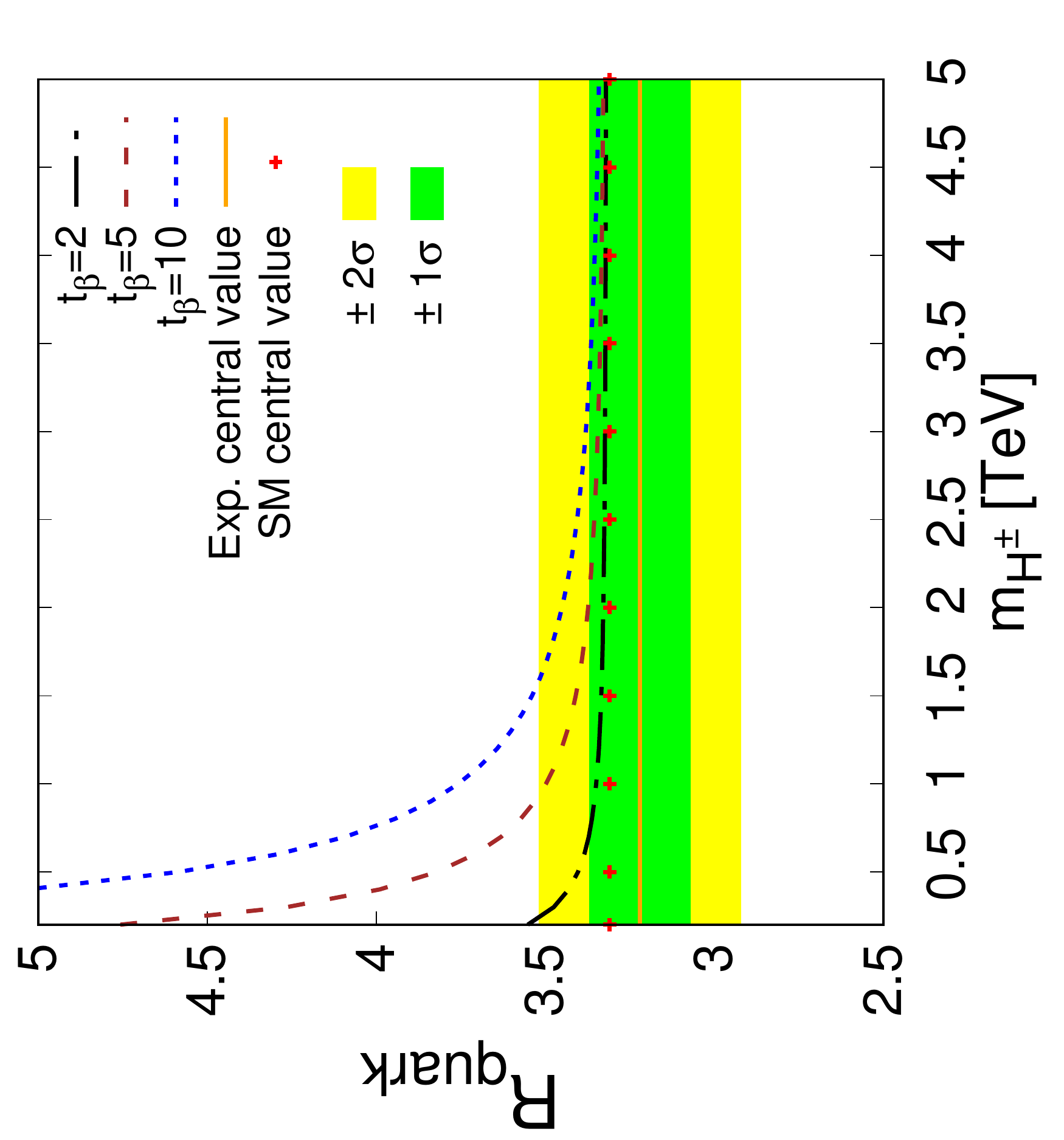}}
 \caption{$R_{quark}$ at NLO in QCD as a function of the charged scalar boson mass for $t_{\beta}=$2, 5, 10. Solid line represents the experimental central value while red crosses indicate the theoretical SM central value. Green and yellow bands stand for 1$\sigma$ and 2$\sigma$, respectively. $R_{quark}$ is defined in the main text.  \label{mCH}   }
\end{figure}

In summary, table \ref{ParValues} shows the values of the 2HDM-III parameters involved in the subsequent calculations.

\begin{table}[!ht]
\caption{Values of the parameters used in the calculations. \label{ParValues}}
\begin{centering}
\begin{tabular}{cc}
\hline 
Parameter & Values\tabularnewline
\hline 
\hline 
$c_{\alpha\beta}$ & 0.05\tabularnewline
\hline 
$t_{\beta}$ & 0.1-8\tabularnewline
\hline 
$\chi_{\tau\mu}$ &1\tabularnewline
\hline 
$m_H=m_A$ & 800 GeV \tabularnewline
\hline 
\end{tabular}
\par\end{centering}
\end{table}

\section{Search for the 
	\texorpdfstring{$h\to\tau\mu$}{h-tau-mu}
	decay at future hadron colliders}
\label{SeccionIV}
\begin{figure}[!ht]
\centering
\includegraphics[scale = 0.35, angle=270]{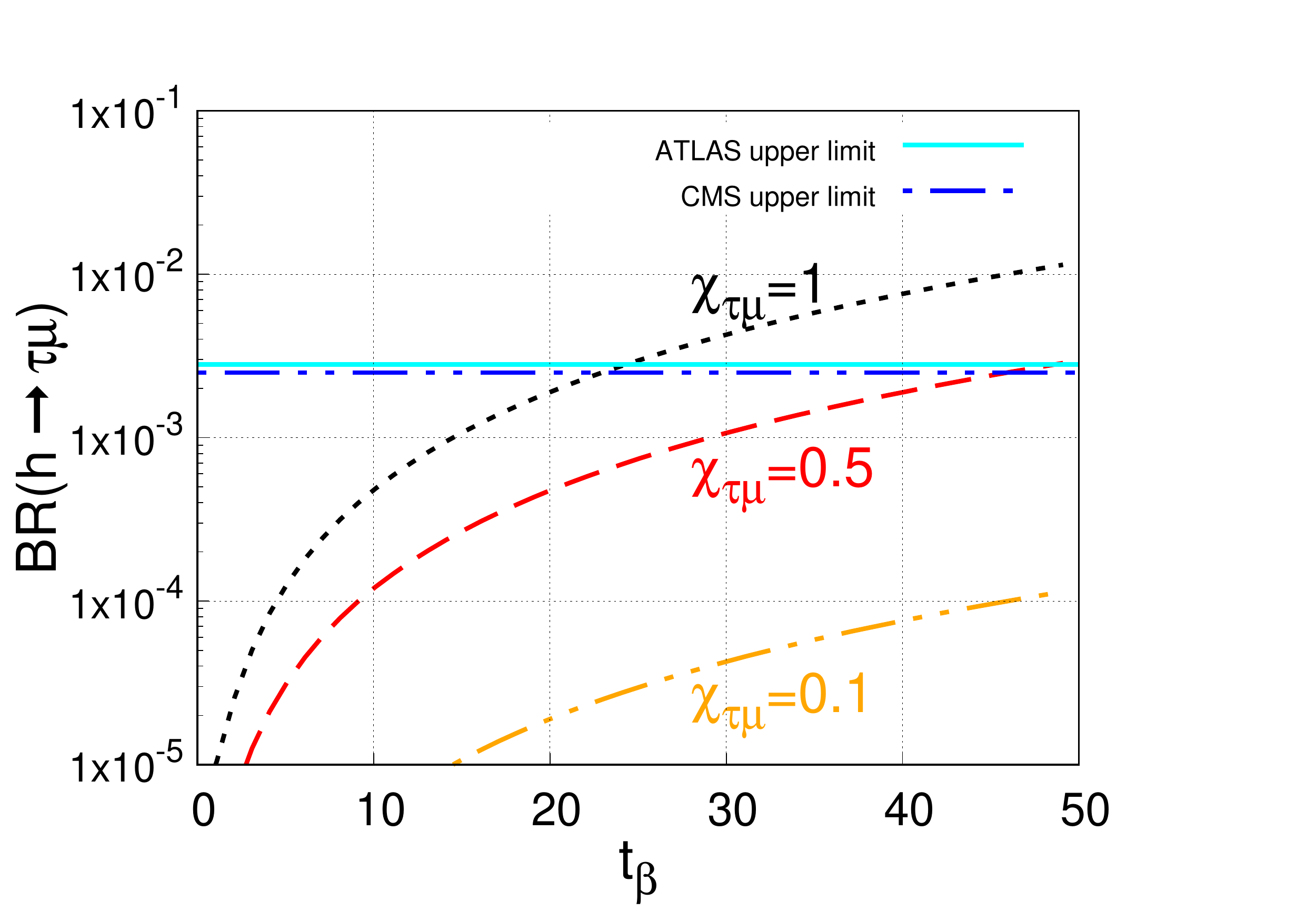}
\caption{Branching ratio of the $h\to\tau\mu$ decay as a function of $t_{\beta}$ for $\chi_{\tau\mu}=0.1,\,0.5,\,1$. The horizontal line represents the upper limit on $\mathcal{BR}(h\to\tau\mu)$.\label{BRhtaumu}}
	\end{figure}

We are interested in a possible evidence for the $h\to\tau\mu$ decay at future hadron collider.  Thus, in this section we analyze the LFV process of the Higgs boson decaying into a $\tau\mu$ pair and its production at future hadron colliders via the gluon fusion mechanism.
We first analyze the behavior of the branching ratio of the $h\to\tau\mu$ decay as a function of $t_{\beta}$ for $\chi_{\tau\mu}=0.1,\,0.5,\,1$ and $c_{\alpha\beta}=0.05$. Figure \ref{BRhtaumu} shows the $\mathcal{BR}(h\to\tau\mu)$ as a function of $t_{\beta}$ including the upper limit on $\mathcal{BR}(h\to\tau\mu)$ reported by CMS and ATLAS collaborations \cite{Sirunyan:2017xzt, ATLAS:2019icc}.
	
We analyze three scenarios that correspond to each of the future hadron colliders, namely:

\begin{itemize}
\item Scenario A (\textbf{SA}): HL-LHC at a center-of-mass energy of 14 TeV and integrated luminosities in the interval 0.3-3 ab$^{-1}$,
\item Scenario B (\textbf{SB}): HE-LHC at a center-of-mass energy of 27 TeV and integrated luminosities in the range 0.3-12 ab$^{-1}$,
\item Scenario C (\textbf{SC}): FCC-hh at a center-of-mass energy of 100 TeV and integrated luminosities from 10 to 30 ab$^{-1}$.
\end{itemize}

\subsection{Number of signal and background events}

Once the free model parameters were constrained in section \ref{SeccionIII}, we now turn to evaluate the number of events produced of the signature $gg\to h\to\tau\mu$. 

\begin{figure}[!ht]
\centering
\subfigure[ ]{\includegraphics[scale = 0.19, angle=270]{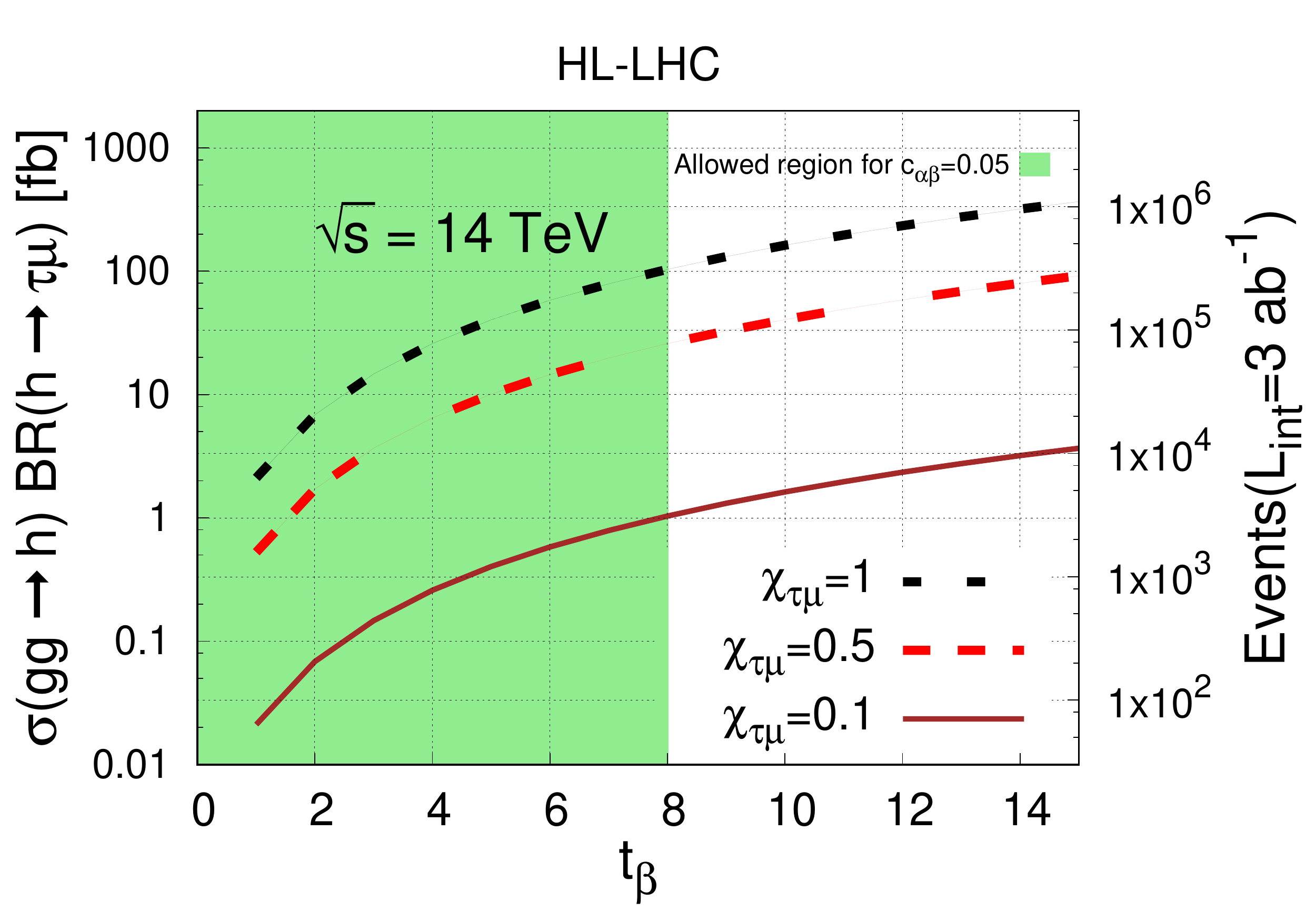}}
\subfigure[ ]{\includegraphics[scale = 0.19, angle=270]{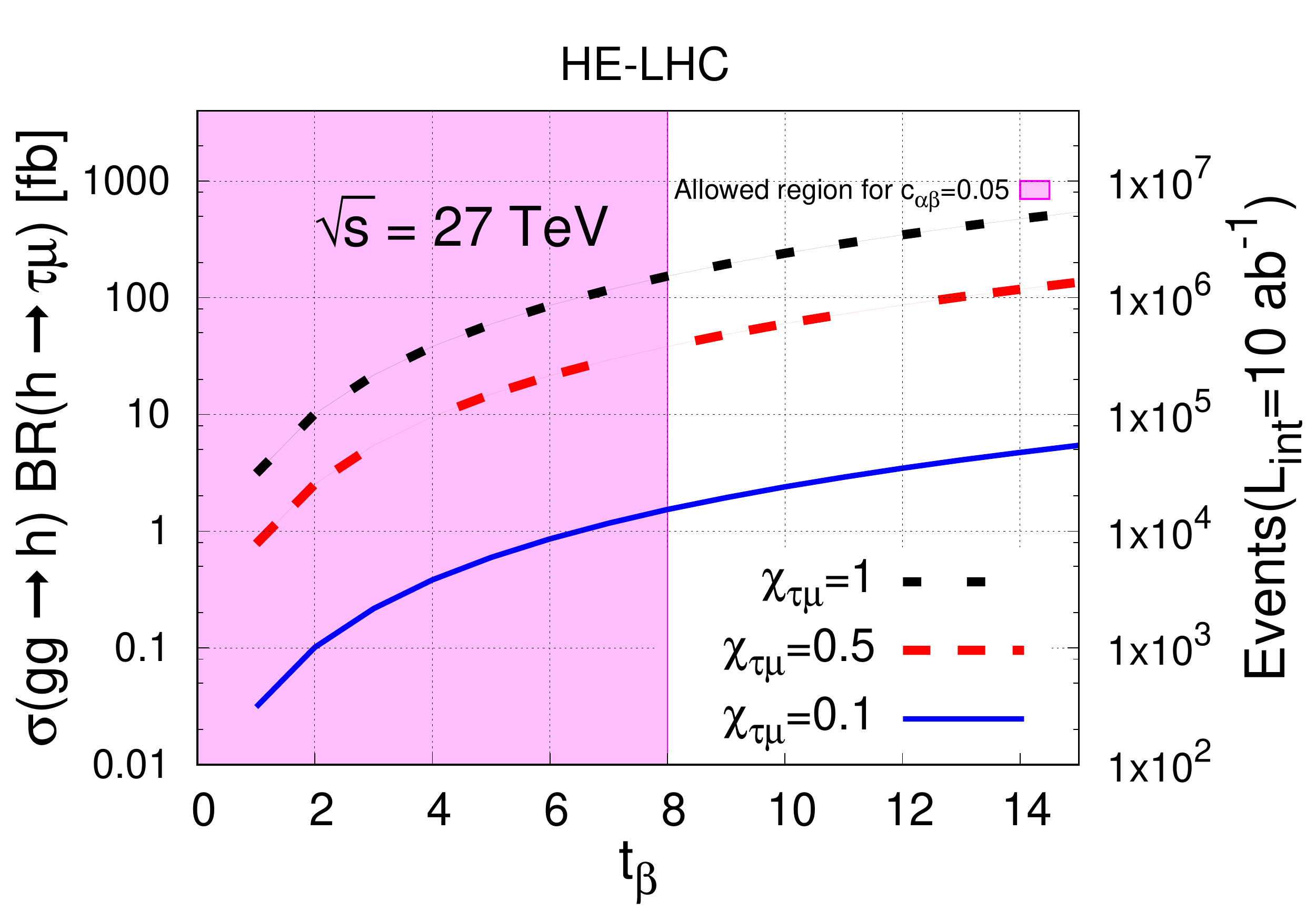}}
\subfigure[ ]{\includegraphics[scale = 0.19, angle=270]{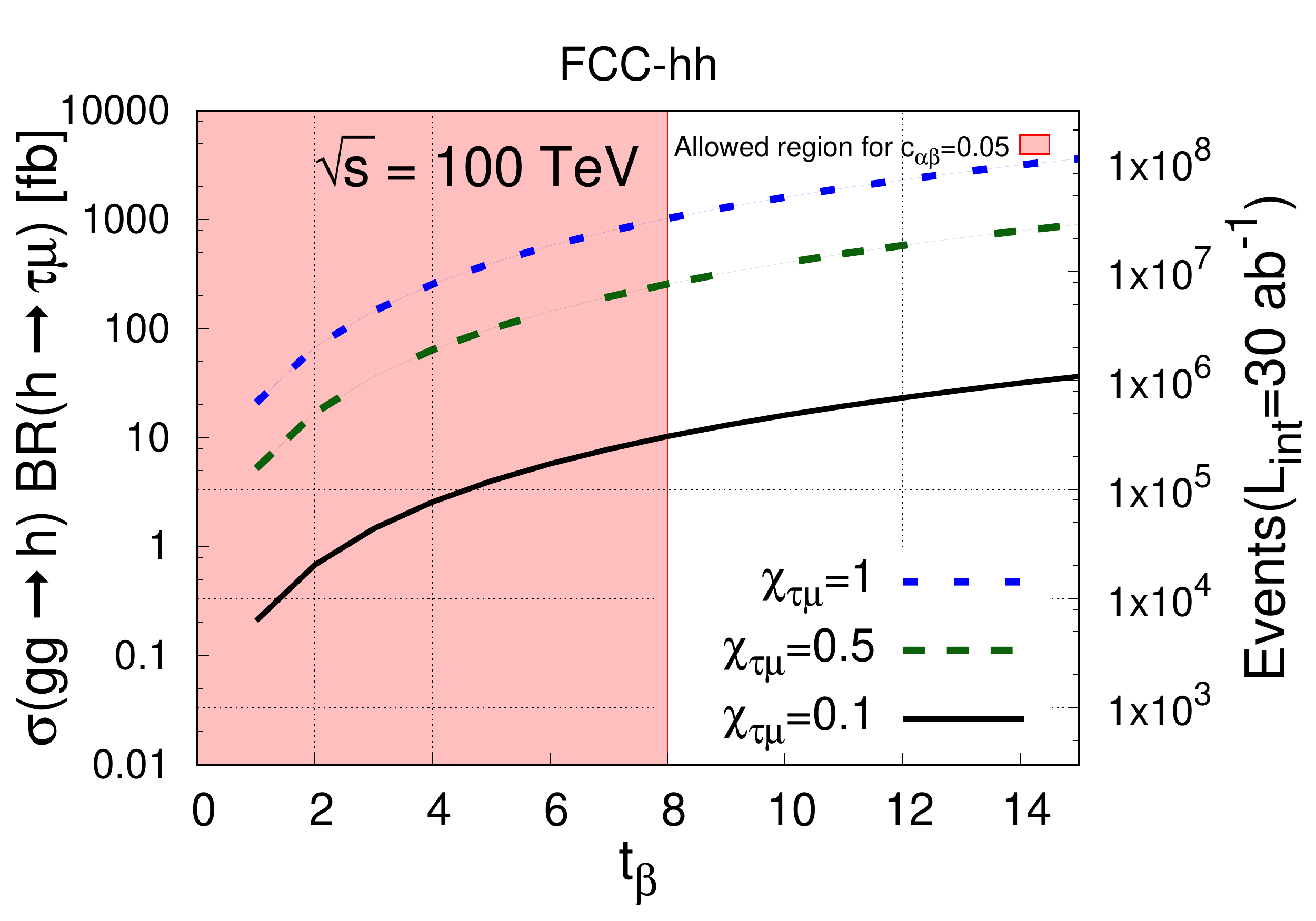}}
\caption{(a) Scenario \textbf{SA}, (b) Scenario \textbf{SB}, (c) Scenario \textbf{SC}. Left axis: $\sigma(gg\to h)\mathcal{BR}(h\to\tau\mu)$ as a function of $t_{\beta}$ for $\chi_{\tau\mu}$=0.1, 0.5, 1. Right axis: Events-$t_{\beta}$ plane. The dark area corresponds to the allowed region. See table \ref{ParValues}. \label{Events}}
\end{figure}

In figure \ref{Events} we present the $\sigma(gg\to h)\mathcal{BR}(h\to\tau\mu)$ as a function of $t_{\beta}$ (left axis) and the Events-$t_{\beta}$ plane (right axis) for scenarios $\textbf{SA}$, $\textbf{SB}$ and $\textbf{SC}$. In all figures, the dark area represents the consistent region with allowed parameter space found in section \ref{SeccionIII} (see table \ref{ParValues}). We observe that the maximum signal number of events ($\mathcal{N}_{\mathcal{S}}^{\textbf{SX}}$) produced are of the order of $\mathcal{N}_{\mathcal{S}}^{\textbf{SA}}=\mathcal{O}(10^5)$, $\mathcal{N}_{\mathcal{S}}^{\textbf{SB}}=\mathcal{O}(10^6)$, $\mathcal{N}_{\mathcal{S}}^{\textbf{SC}}=\mathcal{O}(10^7)$, by considering $t_{\beta}=8$ and $\chi_{\tau\mu}=1$. Where we consider the most up-to-date constrains reported by LHC, in which a value for $t_{\beta}$ of up to 8 is allowed for $c_{\alpha\beta}=0.05$ (see figure \ref{2d}). 

\subsection{Monte Carlo analysis}
We will now analyze the signature of the decay $h\to\tau\mu$, with $\tau\mu=\tau^-\mu^++\tau^+\mu^-$ and its potential SM background. The ATLAS and CMS collaborations \cite{ATLAShtaumu, CMShtaumu} searched two $\tau$ decay channels: electron decay $\tau\to e\nu_{\tau}\nu_{e}$ and hadron decay $\tau_h\mu$. In our analysis, we will concentrate on the electron decay. As far as our computation scheme is concerned, we first implement the relevant Feynman rules via $\texttt{LanHEP}$ \cite{Semenov:2014rea} for $\texttt{MadGraph5}$ \cite{MadGraphNLO}, later it is interfaced with \texttt{Pythia8} \cite{Sjostrand:2008vc} and \texttt{Delphes 3} \cite{delphes} for detector simulations. Subsequently, we generate 10$^5$ signal and background events, the last ones at NLO in QCD. We used $\texttt{CT10}$ parton distribution functions \cite{PDFs}.
\subsection*{Signal and SM background processes}
The signal and background processes are as following:
\begin{itemize}
\item \textbf{SIGNAL:} The signal is $gg\to h\to\tau\mu\to e\nu_{\tau}\nu_{e}\mu$. The electron channel must contain exactly two opposite-charged leptons, namely, one electron and one muon. Therefore, we search for the final state $e\mu$ plus missing energy due to neutrinos not detected.
\item \textbf{BACKGROUND:} The main SM background arises from: 
\begin{enumerate}
\item Drell-Yan process, followed by the decay $Z\to\tau\tau\to e\nu_{\tau}\nu_{e}\mu\nu_{\tau}\nu_{\mu}$.
\item $WW$ production with subsequent decays $W\to e\nu_{e}$ and $W\to \mu\nu_{\mu}$.
\item $ZZ$ production, later decaying into $Z\to\tau\tau\to e\nu_{\tau}\nu_{e}\mu\nu_{\tau}\nu_{\mu}$ and $Z\to\nu\nu$.
\end{enumerate}
\end{itemize}
\subsection*{Signal significance}
The main kinematic cuts to isolate the signal are the collinear and transverse mass defined as following:
\begin{equation}
m_{\text{col}}(e\,\mu)=\frac{m_{\text{inv}}(e\,\mu)}{\sqrt{x}},\, \text{with}\,x=\frac{|\vec{P}_T^e|}{|\vec{P}_T^e|+\vec{E}_T^{\text{miss}}\cdot \vec{P}_T^e}
\end{equation}
and
\begin{equation}
M_T^{\ell}=\sqrt{2|\vec{P}_T^{\ell}||\vec{E}_T^{\text{miss}}|(1-\cos\Delta\phi_{\vec{P}_T^{\ell}-\vec{E}_T^{\text{miss}}})}.
\end{equation}

\begin{figure}[!ht]
\centering
\subfigure[ ]{\includegraphics[scale=0.28]{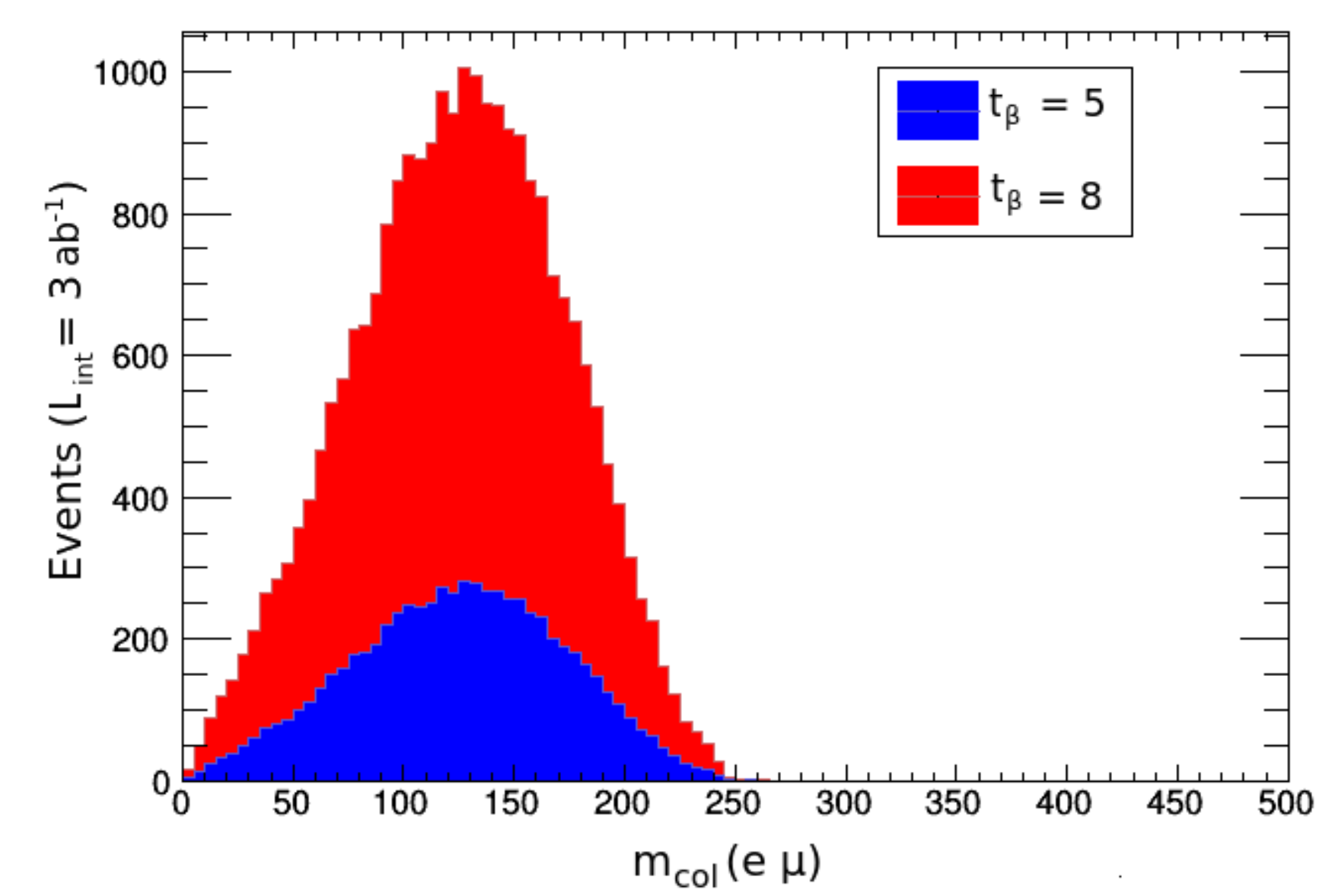}}
\subfigure[ ]{\includegraphics[scale=0.28]{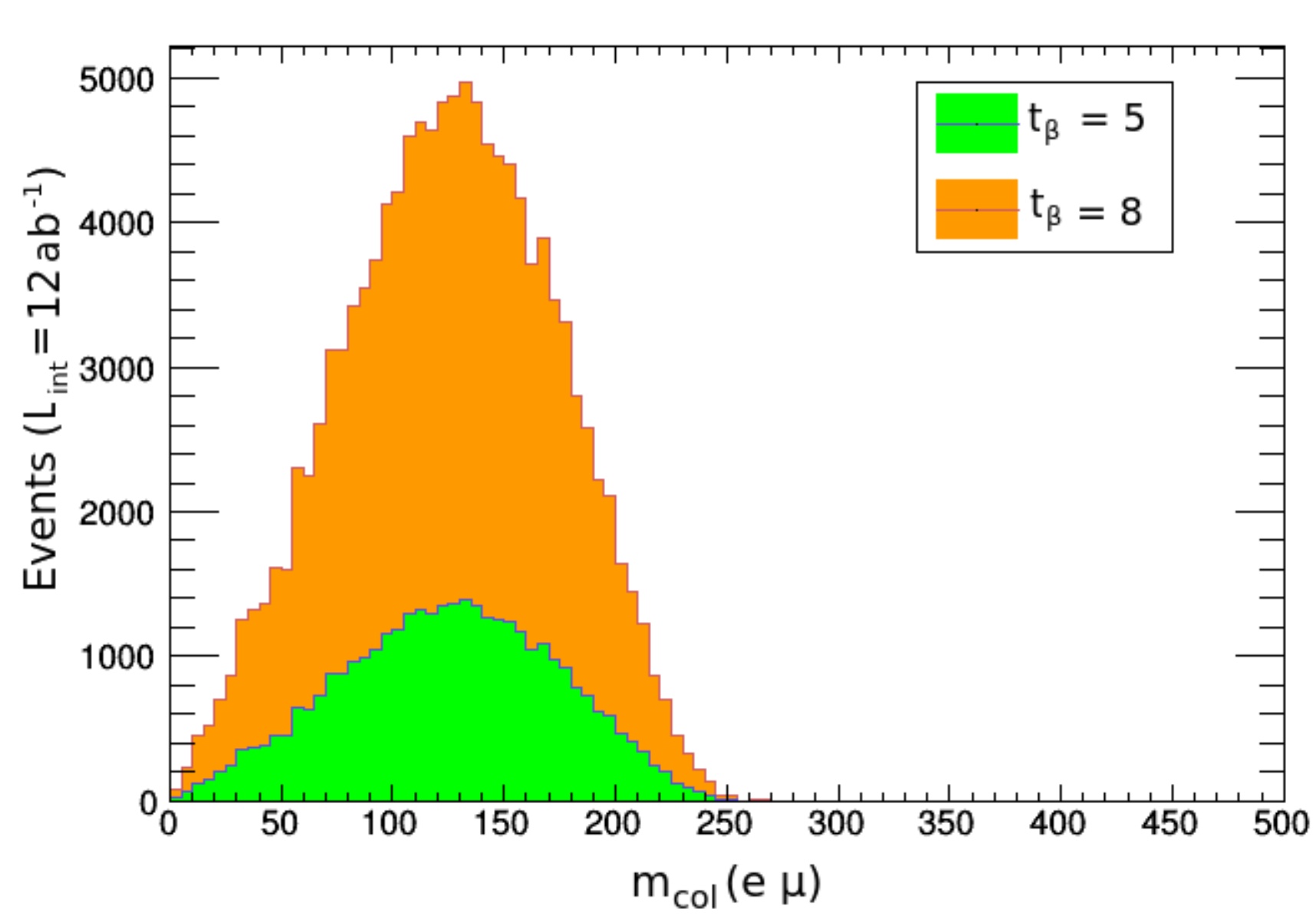}}
\subfigure[ ]{\includegraphics[scale=0.28]{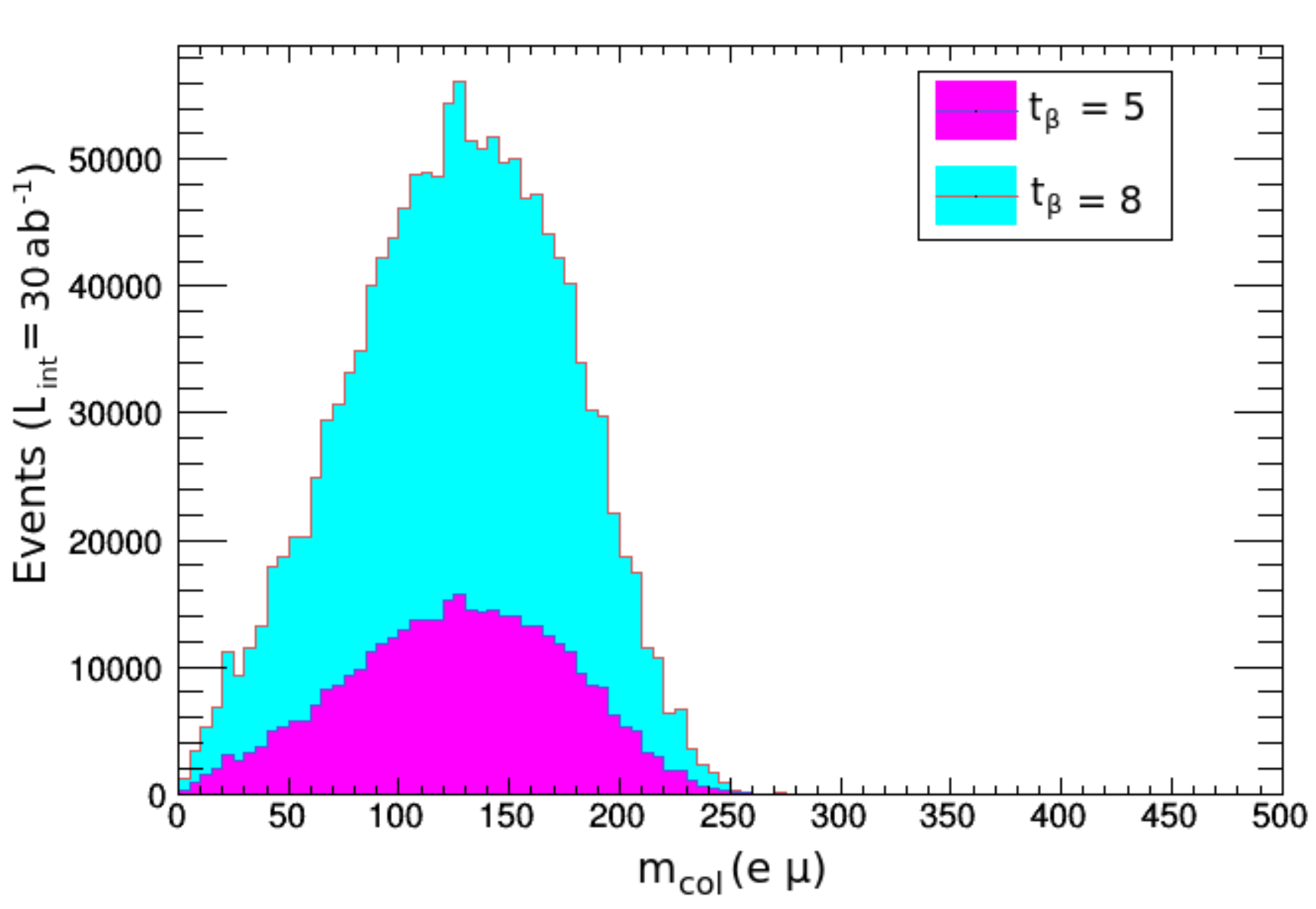}}
 \caption{Distribution of the collinear mass versus number of signal events for scenarios (a) $\textbf{SA}$, (b) $\textbf{SB}$ and (c) $\textbf{SC}$. \label{MCOL}}
	\end{figure}

In figure \ref{MCOL} we show the distribution of collinear mass versus number of signal events for the scenarios (a) $\textbf{SA}$, (b) $\textbf{SB}$ and (c) $\textbf{SC}$ with integrated luminosities of 3, 12 and 30 ab$^{-1}$, respectively. In all scenarios we consider $t_{\beta}$= 5, 8.
We use the package $\texttt{MadAnalysis5}$ \cite{MadAnalysis} to analyze the kinematic distributions. 	
Additional cuts applied both signal and background \cite{ATLAShtaumu, CMShtaumu} are shown in Table \ref{KinematicCutsSA} for scenario $\textbf{SA}$. The kinematic cuts associated to scenarios $\textbf{SB}$ and $\textbf{SC}$ are available electronically in \cite{web_cuts}. We also display the event number of the signal ($\mathcal{N}_S$) and background ($\mathcal{N}_B$) once the kinematic cuts were applied. The signal significance considered is defined as the ratio $\mathcal{N}_S/\sqrt{\mathcal{N}_S+\mathcal{N}_B}$. The efficiency of the cuts for the signal and background are: $\epsilon_S\approx 0.13$ and $\epsilon_B\approx 0.014$, respectively.

\begin{figure}[!ht]
\centering
\subfigure[ ]{\includegraphics[scale = 0.24, angle=270]{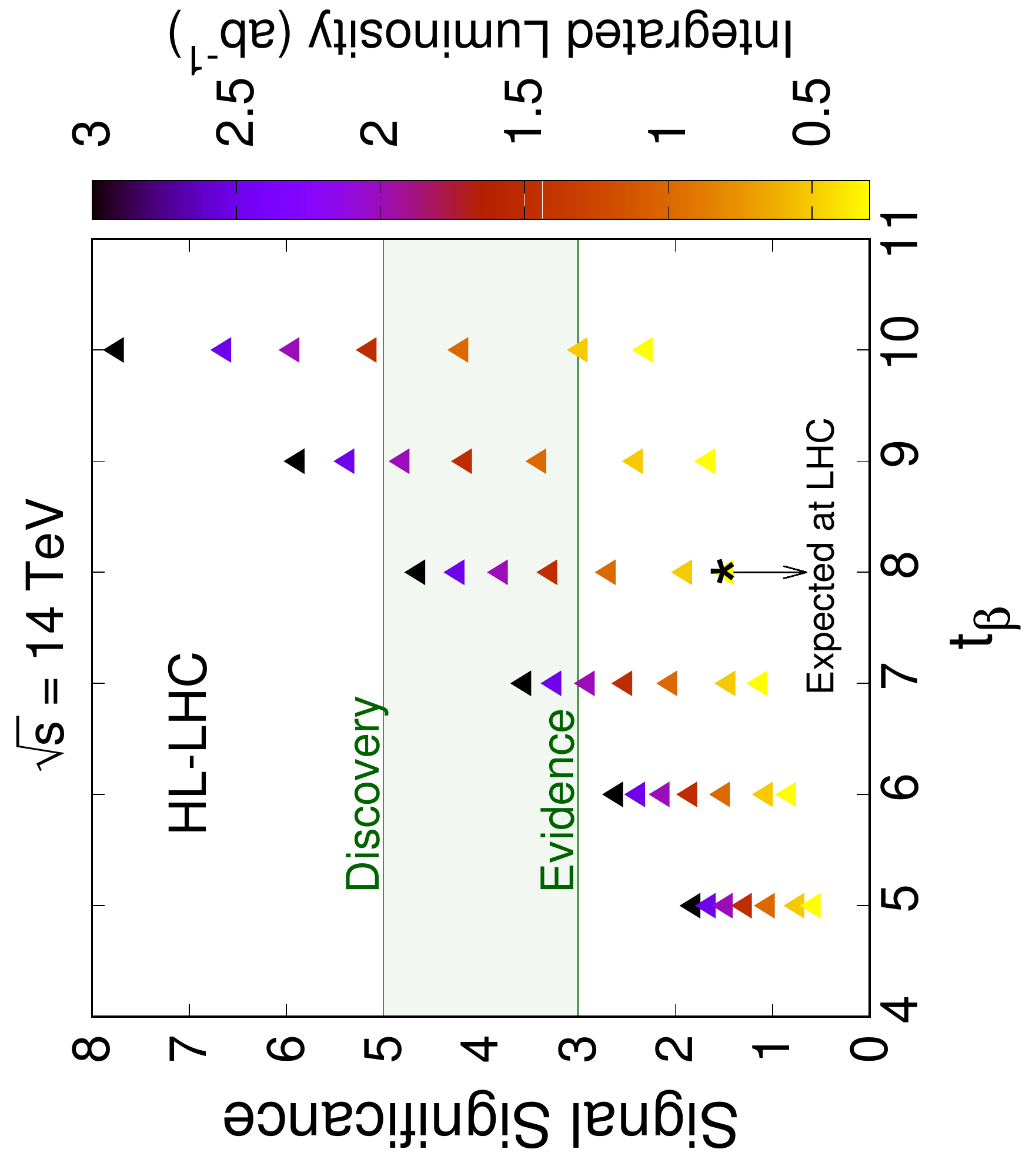}}
\subfigure[ ]{\includegraphics[scale = 0.24, angle=270]{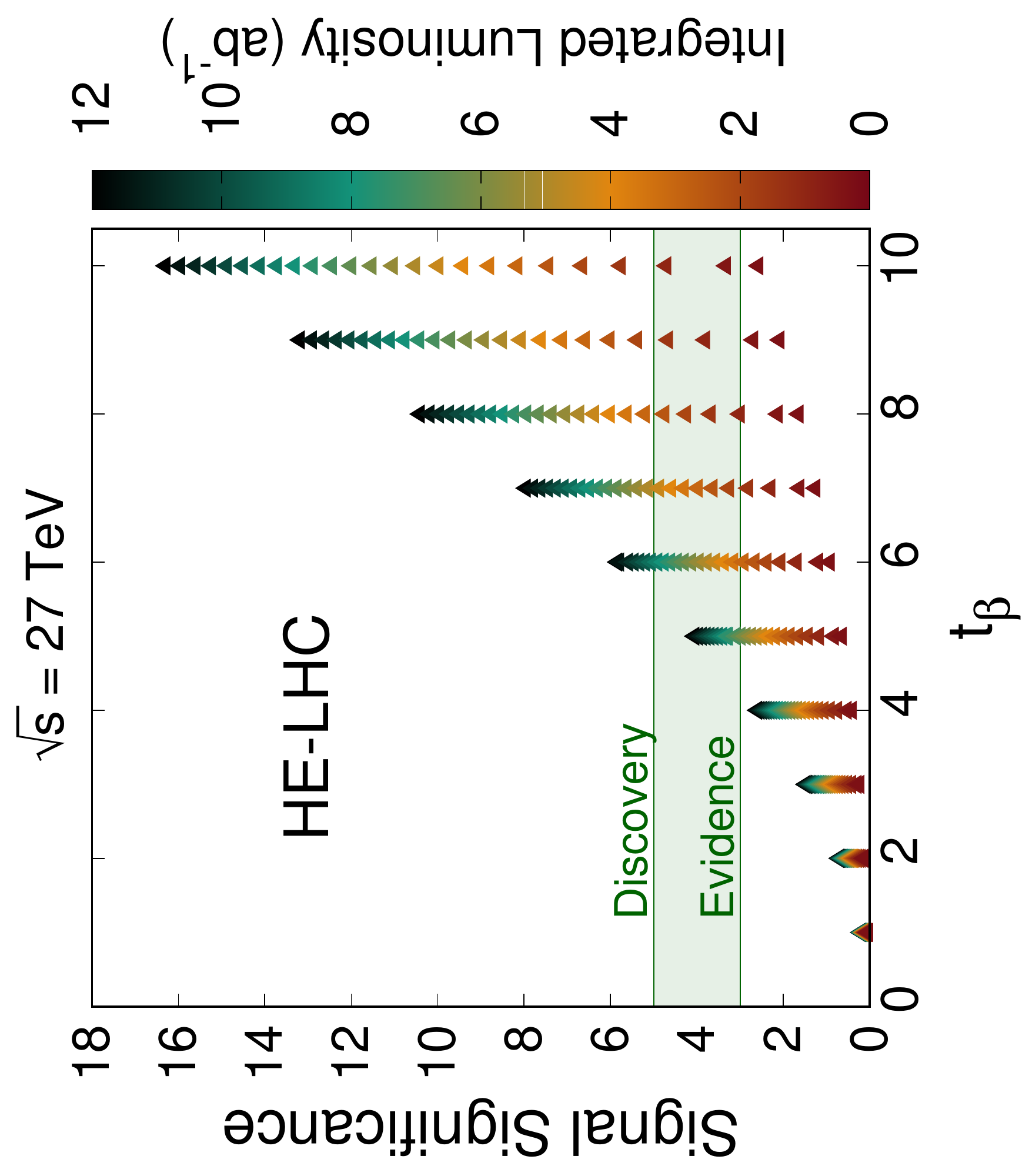}}
\subfigure[ ]{\includegraphics[scale = 0.24, angle=270]{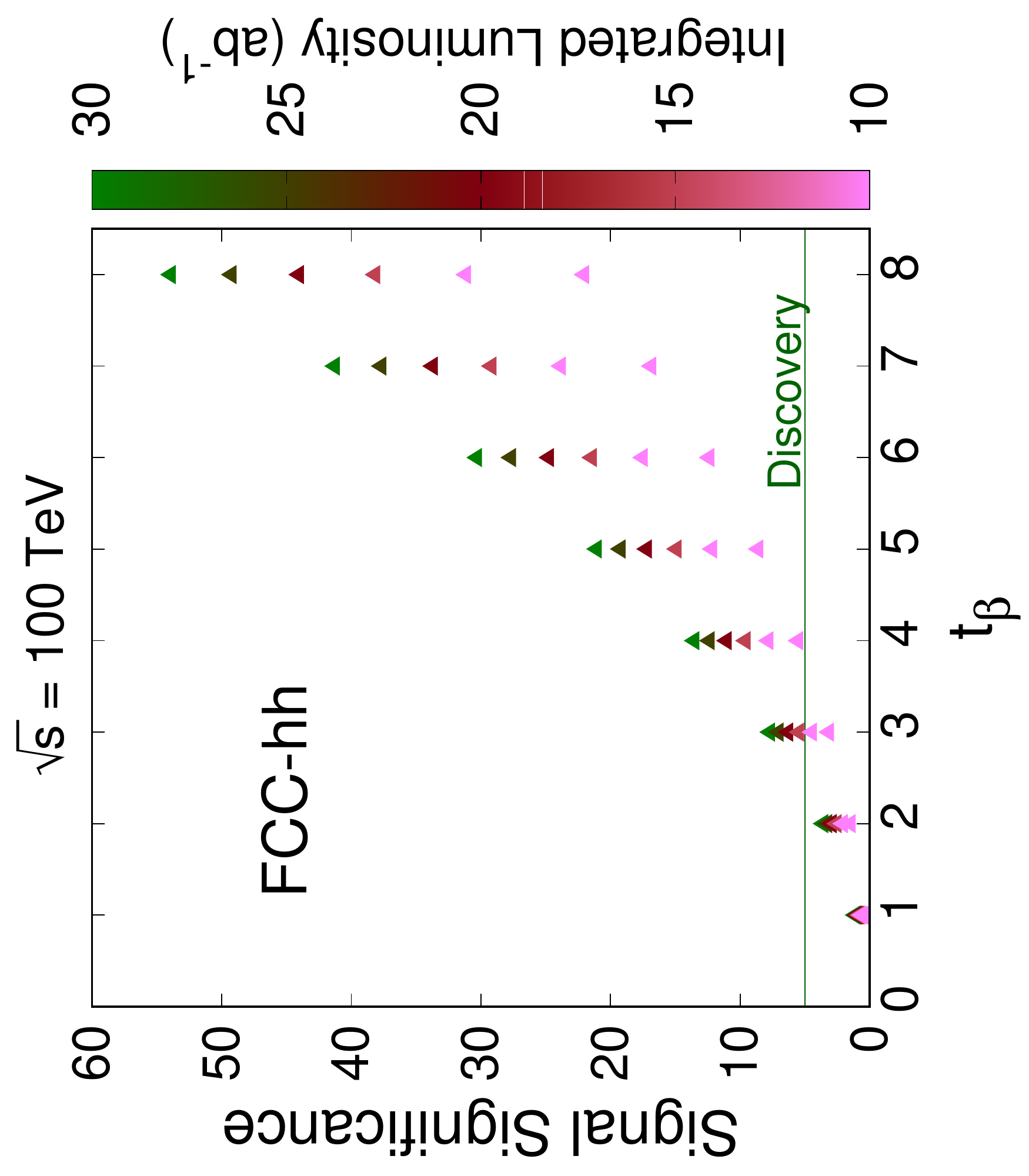}}
\caption{Signal significance as a function of $t_{\beta}$ and integrated luminosities associated to each scenario: (a) \textbf{SA}, (b) \textbf{SB} and (c) \textbf{SC}. \label{Significance_SABC}}
	\end{figure}

\begin{table}[!ht]
\caption{Kinematic cuts applied to the
signal and main SM background for scenario $\mathbf{SA}$, i.e, at HL-LHC with a center-of-mass
energy $\sqrt{s}=14$ TeV and $\mathcal{L}_{\text{int}}=3$ ab$^{-1}$ for $t_{\beta}=8$.}\label{KinematicCutsSA}
\begin{centering}
\begin{tabular}{c c c c c}
\hline 
Cut number & Cut & $\mathcal{N}_S$ & $\mathcal{N}_B$ & $\mathcal{N}_S/\sqrt{\mathcal{N}_S+\mathcal{N}_B}$
\tabularnewline
\hline 
\hline 
 & Initial (no cuts) & $57665$ & $200089020$ & $4.08$\tabularnewline
\hline 
$1$ & $|\eta^{e}|<2.3$ & $25282$ & $132346436$ & $2.1975$\tabularnewline
\hline 
$2$ & $|\eta^{\mu}|<2.1$ & $16378$ & $106936728$ & $1.5837$\tabularnewline
\hline 
$3$ & $0.1<\Delta R(e,\,\mu)$ & $16355$ & $106801230$ & $1.5825$\tabularnewline
\hline 
$4$ & $10<p_{T}(e)$ & $15533$ & $38846174$ & $2.4817$\tabularnewline
\hline 
$5$ & $20<p_{T}(\mu)$ & $12119$ & $20357367$ & $2.6852$\tabularnewline
\hline 
$6$ & $10<\text{MET}$  & $11185.9$ & $20086662$ & $2.4952$\tabularnewline
\hline 
$7$ & $100<m_{\text{col}}(e,\,\mu)<150$  & $9645.1$ & $9330510$ & $3.1560$\tabularnewline
\hline 
$8$ & $25<M_{T}(e)$ & $8669.4$ & $4827617$ & $3.942$\tabularnewline
\hline 
$9$ & $15<M_{T}(\mu)$ & $7869$ & $2867711$ & $4.6404$\tabularnewline
\hline 
\end{tabular}
\par\end{centering}
\end{table}
We find that at the LHC is not possible to claim for evidence of the decay $h\to\tau\mu$ achieving a signal significance about $1.46\sigma$ by considering its final integrated luminosity, 300 fb$^{-1}$. More promising results arise at HL-LHC in which a prediction of about $4.6\sigma$, once an integrated luminosity of 3 ab$^{-1}$ and $t_{\beta}=8$ are achieved. Meanwhile, at HE-LHC (FCC-hh) a potential discovery could be claimed with a signal significance of around $5.04\sigma$ ($\sim 5.43\sigma$) for an integrated luminosity of 9 ab$^{-1}$ and $t_{\beta}=6$ (15 ab$^{-1}$ and $t_{\beta}=3$).  
To illustrate the above, in figure \ref{Significance_SABC} we present the signal significance as a function of $t_{\beta}$ for integrated luminosities associated with each scenario, namely:
\begin{itemize}
\item \textbf{SA}: from 0.3 ab$^{-1}$ at 3 ab$^{-1}$ for the HL-LHC,
\item \textbf{SB}: from 3 ab$^{-1}$ at 12 ab$^{-1}$ for the HE-LHC,
\item \textbf{SC}: from 10 ab$^{-1}$ at 30 ab$^{-1}$ for the FCC-hh.
\end{itemize}

\begin{figure}[!ht]
\centering
\includegraphics[scale = 0.35]{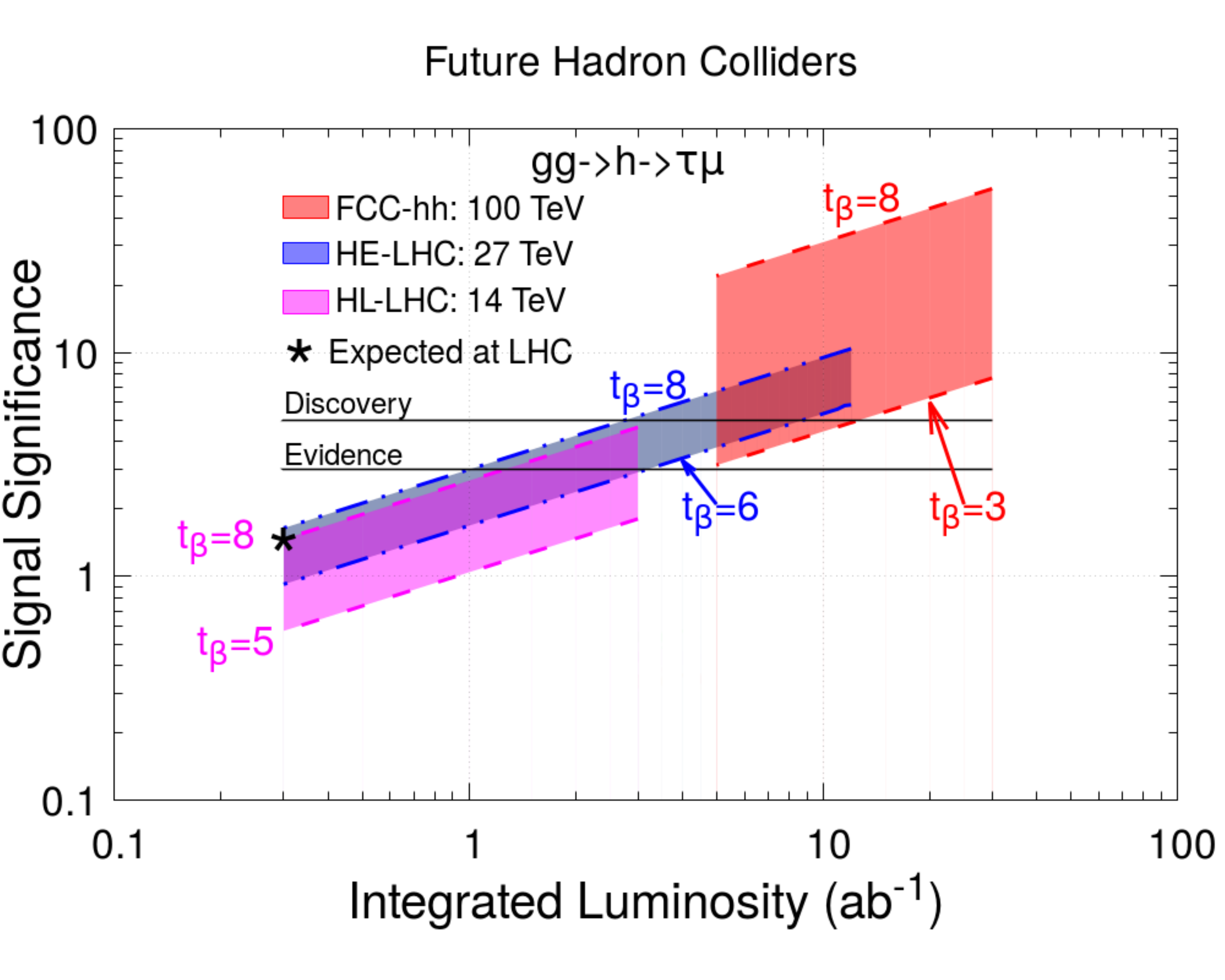}
\caption{Signal significance as a function of the integrated luminosity for representative values of $t_{\beta}$. \label{ALLscenarios}}
	\end{figure}

Finally, we present in figure \ref{ALLscenarios} an overview of the signal significance as a function of the integrated luminosity for representative values of $t_{\beta}$.

\newpage
\section{Conclusions}
\label{SeccionV}

	In this article we have studied the LFV decay $h\to\tau\mu$ within the context of the 2HDM type III and we analyze its possible detectability at future super hadron colliders, namely, HL-LHC, HE-LHC and the FCC-hh.
	
	We find the allowed model parameter space by considering the most up-to-date experimental measurements and later is used to evaluate the Higgs boson production cross section via the gluon fusion mechanism and the branching ratio of the $h\to\tau\mu$ decay. 
	
A Monte Carlo analysis of the signal and its potential SM background was realized. We find that the closest evidence could arise at the HL-LHC with a prediction of the order of 4.66$\sigma$ for an integrated luminosity of 3 ab$^{-1}$ and $\tan\beta=8$. On the other hand, a potential discovery could be claimed at the HE-LHC (FCC-hh) with a signal significance about 5.046$\sigma$ (5.43$\sigma$) for an integrated luminosity of 3 ab$^{-1}$ and $\tan\beta=8$ (5 ab$^{-1}$ and $\tan\beta=4$).

If the decay considered in this research is observed in a future super hadron collider, then it will be a clear signal of physics BSM.

%
%

\section*{Acknowledgments}
Marco Antonio Arroyo Ure\~na especially thanks to \emph{PROGRAMA DE BECAS POSDOCTORALES DGAPA-UNAM} for postdoctoral funding. 
This work was supported by projects \emph{Pro\-gra\-ma de A\-po\-yo a Proyectos de Investigaci\'on e Innovaci\'on
Tecnol\'ogica} (PAPIIT) with registration codes IA107118 and IN115319 in \emph{Direcci\'on General de Asuntos de Personal
Acad\'emico de Universidad Nacional Aut\'onoma de M\'exico} (DGAPA-UNAM), and \emph{Programa Interno de Apoyo para
Proyectos de Investigaci\'on} (PIAPI) with registration code PIAPI1844 in FES-Cuautitl\'an UNAM and \emph{Sistema Nacional de Investigadores} (SNI) of the \emph{Consejo Nacional de Ciencia y Tecnolog\'ia} (CO\-NA\-CYT) in M\'exico. Also we would like to thank CO\-NA\-CYT for the support of the author T. A. Valencia-P\'erez  with a doctoral grant and thankfully acknowledge computer resources, technical advise and support provided by Laboratorio Nacional de Superc\'omputo del Sureste de M\'exico.
%

\appendix

\section{Complementary formulas used in the analysis of the model parameter space}\label{FormulaDecays}

In this Appendix we present the analytical expressions in order to obtain the constraints on both diagonal and LFV couplings as is shown in Figure \ref{tb_cab}.

\subsection{SM-like Higgs boson into 
	\texorpdfstring{$f_i\bar{f}_j$}{fifj}
}

We first start with the expression for the width decay of SM-like Higgs boson into fermion pair, which is given by:
\begin{equation} \label{widDecHiggsfifj}
\Gamma(h\to f_i\bar{f}_j)=\frac{g^2_{hf_i\bar{f}_j}N_c m_{h}}{128\pi}\left(4-\left(\sqrt{\lambda_{f_i}}+\sqrt{\lambda_{f_j}}\right)^2\right)^{3/2}
  \left(4-\left( \sqrt{   \lambda_{f_i}}-\sqrt{\lambda_{f_j}} \right)^2  \right)^{1/2}  ,
\end{equation}
where $\lambda_{f_k}=4m_{f_k}^2/m_h^2$, with $k=i,\,j$; $N_c$ is the color number. In our case $g_{h\tau\mu}=\frac{c_{\alpha\beta}t_{\beta}}{\sqrt{2}s_{\beta}}\tilde{Y}_{\tau\mu}$ with $\tilde{Y}_{\tau\mu}=\frac{\sqrt{m_{\tau}m_{\mu}}}{\upsilon} \chi_{\tau\mu}$. We set $\chi_{\tau\mu}=1$. 

\subsection{Tau decays 
	\texorpdfstring{$\tau\to\mu\gamma$}{tau->mu gamma}
	and 
	\texorpdfstring{$\tau\to\mu\bar\mu \mu$}{tau->mu mu mu}
}

\begin{figure}[!ht]
  \centering
  \includegraphics[width=10cm]{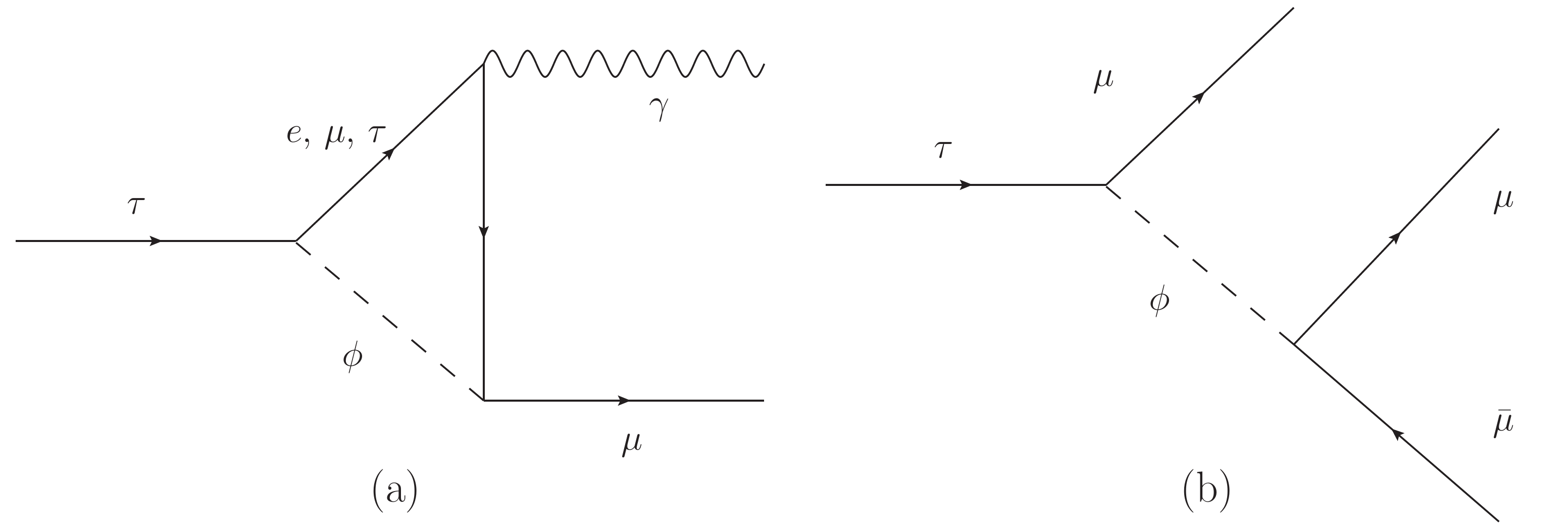}
  \caption{Feynman diagrams that contribute to  (a) $\tau\to\mu\gamma$ and (b) $\tau\to\mu\bar\mu \mu$ decays with exchange of a scalar boson $\phi$. We omit both the bubble diagrams for the LFV decay $\tau\to\mu\gamma$, because only serve to cancel the ultraviolet divergences.}\label{LFVScalarDecays}
\end{figure}

As far as the $\tau\to\mu\gamma$ decay is concerned, it arises at the one-loop level and receives contributions of $\phi=h,\,H,\,A$. Feynman diagrams for this process are displayed in figure \ref{LFVScalarDecays}(a). The decay width is given by:
\begin{equation}
\Gamma(\tau\to\mu\gamma)=\frac{\alpha m_{\tau}^{5}}{64\pi^{4}}\left(|A_{S}|^{2}+|A_{P}|^{2}\right),
\end{equation}
where the $A_S$ and $A_P$ coefficients indicate the contribution from $A$ and $H$, respectively. In the limit of $g_{\phi\tau\tau}\gg g_{\phi \mu\mu} \gg g_{\phi ee}$ and $m_\tau\gg m_\mu\gg m_e$,
 they can be approximated as \cite{Harnik:2012pb}
\begin{equation}
A_{S}  = A_P\simeq \sum_{\phi=h,H_F, A_F} \frac{g_{\phi\tau\tau}g_{\phi \mu\tau}}{12m_{\phi}^{2}}\left(3\ln\left(\frac{m_\phi^2}{m_{\tau}^2}\right)-4\right).
\end{equation}
Two-loop contributions can be relevant, their expressions are reported in \cite{Harnik:2012pb}, in our research we consider this contribution. The current experimental limit on the branching ratio is $\mathcal{BR}(\tau\to\mu\gamma)<4.4\times10^{-8}$.

As for the $\tau \to \mu\bar\mu \mu$ decay, it receives contributions from $\phi$ as depicted in  the Feynman diagram of Figure \ref{LFVScalarDecays}(b). The tree-level decay width can  be approximated as

\begin{align}
\Gamma(\tau\to \mu\bar\mu \mu)&\simeq\frac{m_\tau^5}{256 \pi^3}\Bigg(\frac{S_h^2}{m_h^4}+\frac{S_{H}^2}{m_{H}^4}+\frac{S_{H}^2}{m_{H}^4}+ \frac{2 S_h S_{H}}{m_h^2 m_{H}^2}+\frac{2 S_{A}}{3m_{A}^2}\left(\frac{S_h}{m_h^2}+\frac{S_{H}}{m_{H}}\right)\Bigg),
\end{align}
where $S_\phi=g_{\phi \mu\mu} g_{\phi \mu\tau}$.
The upper bound on the branching ratio is $BR(\tau\to\mu\bar\mu\mu)<2.1\times10^{-8}$ \cite{PDG}.
\subsection{Muon anomalous magnetic dipole moment}
The muon AMDM also receives contributions from $\phi$, which are induced by a triangle diagram similar to the diagram of Figure \ref{LFVScalarDecays}(a) but with two external muons.  The corresponding contribution  can be approximated for $m_\phi\gg m_l$ as \cite{Harnik:2012pb}
\begin{equation}
\label{deltaamu}
\delta a_{\mu}\sim\frac{m_{\mu}}{16\pi^{2}}\sum_{\phi=h,H,A}\sum_{l=\mu,\tau}\frac{m_l g_{\phi \mu l}^2 }{m_{\phi}^{2}}\left(2\ln\left(\frac{m_{\phi}^{2}}{m_l^{2}}\right)-3\right),
\end{equation}
where  one must take into account the NP corrections to the $g_{h\mu\mu}$ coupling   only. If $H$ and $A$ are too heavy, the dominant  NP contribution would arise from the SM Higgs boson.

The discrepancy between the experimental value and the SM theoretical prediction  is 
\begin{equation}
\Delta a_{\mu}=a_{\mu}^{exp}-a_{\mu}^{SM}=(2.88\pm 0.63\pm 0.49)\times 10^{-9}.
\end{equation}
Thus, the requirement that this  discrepancy is accounted for by  Eq. (\ref{deltaamu}) leads to the bound $1.32 \times 10^{-9}\le \Delta a_{\mu}\le 4.44\times 10^{-9}$ with $95\%$ C.L.

\subsection{Decay 
	\texorpdfstring{$B_s^0\to\mu^-\mu^+$}{B->mu-mu+}
}
\begin{figure}[!ht]
  \centering
  \includegraphics[width=5cm]{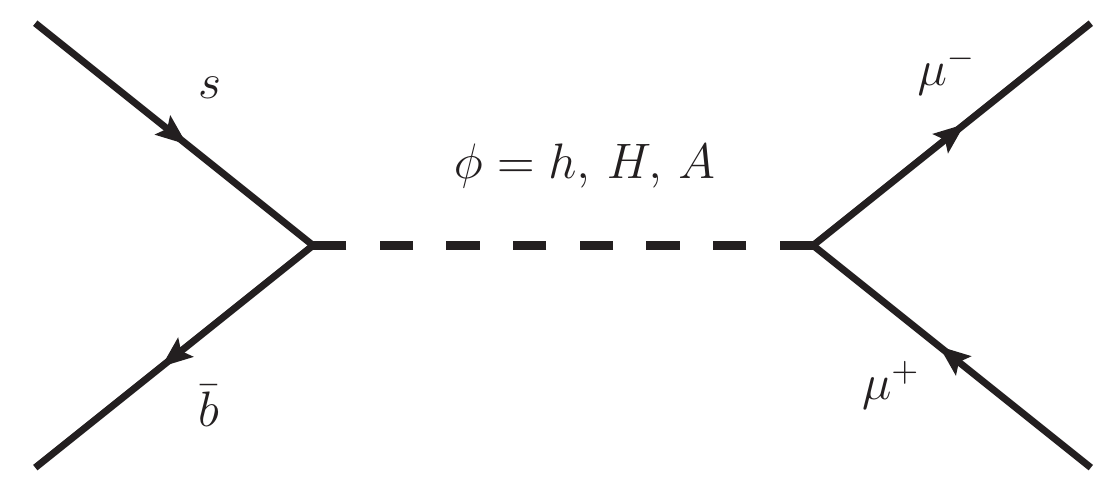}
  \caption{Feynman diagram for the decay $B_s^0\to\mu^+\mu^-$.}\label{Bmumu}
\end{figure}

$B_s^0$ meson decay into $\mu^+\mu^-$ pair is both interesting and stringent due to its sensitivity to constrain BSM theories. The SM theoretical prediction is $3.660\times 10^{-9}$ \cite{Beneke:2019slt} while the experimental value is $(3.00\pm0.6^{+0.3}_{-0.2})\times 10^{-9}$ \cite{PDG}. In the context of the THDM-III, the decay $B_s^0\to\mu^+\mu^-$ is mediated by the SM-like Higgs boson, the heavy scalar $H$ and the pseudoscalar $A$ and it arises at tree level. Feynman diagram at the quark level is shown in Figure \ref{Bmumu}.  
The branching ratio for this decay is given by \cite{Crivellin:2013wna}
\begin{align}
\begin{array}{ccl}
\mathcal{BR}\left[B_{s}\rightarrow\mu^{+}\mu^{-}\right] & = & \frac{G_{F}^{4}M_{W}^{4}}{8\pi^{5}}\sqrt{1-4\frac{m_{\mu}^{2}}{M_{B_{s}}^{2}}}M_{B_{s}}f_{B_{s}}^{2}m_{\mu}^{2}\tau_{B_{s}}\\
& \times & \left[\left|\frac{M_{B_{s}}^{2}\left(C_{P}^{bs}-C_{P}^{\prime bs}\right)}{2\left(m_{b}+m_{s}\right)m_{\mu}}-\left(C_{A}^{bs}-C_{A}^{\prime bs}\right)\right|^{2}+\left|\frac{M_{B_{s}}^{2}\left(C_{S}^{bs}-C_{S}^{\prime bs}\right)}{2\left(m_{b}+m_{s}\right)m_{\mu}}\right|^{2}\times\left(1-4\frac{m_{\mu}^{2}}{m_{B_{s}}^{2}}\right)\right]
\end{array}
\end{align}

\begin{align}
\begin{array}{c}
C_{S}^{q_{f}q_{i}}=\frac{\pi^{2}}{2G_{F}^{2}M_{W}^{2}}\sum_{k=1}^{3}\frac{1}{m_{H_{k}^{0}}^{2}}\left(\Gamma_{l_{B}l_{A}}^{LRH_{k}^{0}}+\Gamma_{l_{B}l_{A}}^{RLH_{k}^{0}}\right)\Gamma_{q_{f}q_{i}}^{RLH_{k}^{0}}\\
C_{P}^{q_{f}q_{i}}=\frac{\pi^{2}}{2G_{F}^{2}M_{W}^{2}}\sum_{k=1}^{3}\frac{1}{m_{H_{k}^{0}}^{2}}\left(\Gamma_{l_{B}l_{A}}^{LRH_{k}^{0}}-\Gamma_{l_{B}l_{A}}^{RLH_{k}^{0}}\right)\Gamma_{q_{f}q_{i}}^{RLH_{k}^{0}}\\
C_{S}^{\prime q_{f}q_{i}}=\frac{\pi^{2}}{2G_{F}^{2}M_{W}^{2}}\sum_{k=1}^{3}\frac{1}{m_{H_{k}^{0}}^{2}}\left(\Gamma_{l_{B}l_{A}}^{LRH_{k}^{0}}+\Gamma_{l_{B}l_{A}}^{RLH_{k}^{0}}\right)\Gamma_{q_{f}q_{i}}^{LRH_{k}^{0}}\\
C_{P}^{\prime q_{f}q_{i}}=\frac{\pi^{2}}{2G_{F}^{2}M_{W}^{2}}\sum_{k=1}^{3}\frac{1}{m_{H_{k}^{0}}^{2}}\left(\Gamma_{l_{B}l_{A}}^{LRH_{k}^{0}}-\Gamma_{l_{B}l_{A}}^{RLH_{k}^{0}}\right)\Gamma_{q_{f}q_{i}}^{LRH_{k}^{0}}
\end{array}
\end{align}

\end{document}